\documentclass[aps,prl,reprint,superscriptaddress,nofootinbib]{revtex4-2}

\usepackage{amsmath}
\usepackage{amssymb}
\usepackage{graphicx}
\usepackage{bm}
\usepackage{mathtools}
\usepackage[dvipsnames]{xcolor}
\usepackage[colorlinks, linkcolor=BrickRed,anchorcolor=blue,citecolor=blue,urlcolor=blue]{hyperref}

\usepackage{pgffor}
\usepackage{pdfpages}
\makeatletter
\AtBeginDocument{\let\LS@rot\@undefined}
\makeatother

\newcommand{\Tr}{\operatorname{Tr}}
\newcommand{\ii}{\mathrm{i}}
\newcommand{\cH}{C^{\mathrm H}}
\newcommand{\cn}[1]{C^{(#1)}}
\newcommand{\id}{\mathbb{I}}
\newcommand{\norm}[1]{\left\lVert #1\right\rVert}
\newtheorem{theorem}{Theorem}

\begin{document}

\title{Closing a single-copy gap to the Holevo bound with two-copy measurements}

\author{Changhao Li}
\thanks{changhaoli96@gmail.com}
\affiliation{Unitary Foundation, San Francisco, CA, USA}

\date{\today}

\begin{abstract}
The Holevo Cram\'{e}r-Rao bound sets the asymptotically attainable precision of quantum multiparameter estimation, but its attainability with finite quantum state copies remains unresolved in general. While a gap between the single-copy optimum and the Holevo bound persists at any finite copy number for two-parameter estimation models, the higher-parameter case has remained open. Here we disprove universal finite-copy gap persistence beyond two parameters by constructing a three-parameter rank-two model whose Holevo bound is unattainable with one copy but attained exactly with two. We further prove finite-copy persistence for full-rank models with any number of parameters. We then experimentally implement the two-copy collective measurement for the constructed model using superconducting qubits. We observe a two-copy estimation error below the single-copy and separate-measurement bounds determined from state tomography. By introducing controlled depolarization, we show that this precision advantage persists over a finite noise range. Our results connect the attainability of ultimate quantum precision to the structure of the multiparameter estimation model and the resources available for collective measurements.
\end{abstract}

\maketitle

\textit{Introduction}. The central goal of quantum sensing is to estimate physical parameters with the highest precision allowed by quantum mechanics~\cite{Degen2017,Giovannetti2011,Pezze2018,Paris2009}. When several parameters are estimated simultaneously, measurements optimized for different quantities can be incompatible, creating tradeoffs absent in single-parameter sensing~\cite{Szczykulska2016,Albarelli2020perspective,Sidhu2020,Liu2020,Li2022geometric,Hu2024control,Ragy2016}. 
The Holevo Cram\'{e}r-Rao bound incorporates these tradeoffs and sets the ultimate precision limit~\cite{Holevo2011,Yamagata2013,Yang2019,Demkowicz2020,SidhuOuyang2021}, and understanding how to approach it is important for applications such as quantum imaging~\cite{Tsang2016superresolution,Rehacek2017imaging,Parniak2018,deshler2608.19524} and vector magnetometry~\cite{BaumgratzDatta2016,Liu2019vector,Wang2021vector}. 
Attaining this limit is guaranteed under standard regularity conditions when collective measurements on asymptotically many identical copies are allowed~\cite{HayashiMatsumoto2008,KahnGuta2009,Yamagata2013,Yang2019,Tsang2026PRL}. Practical measurements, however, can jointly access only a finite number of copies. Whether a finite collective measurement can ever attain the Holevo bound exactly is therefore a fundamental question~\cite{Matsumoto2002,Conlon2022gap,Horodecki2022}.

Collective measurements on finite ensembles have long been shown to outperform separate measurements in quantum state estimation~\cite{MassarPopescu1995,Bagan2004,Hou2018}, and recent multiparameter studies have analyzed and demonstrated related few-copy advantages~\cite{Conlon2021,Friel2020,Conlon2023collective,Das2025,Parniak2018,Roccia2018}. Such an advantage does not imply attainment of the Holevo bound. A finite-copy protocol may improve precision while remaining strictly above the asymptotic limit. For two-parameter models, the gap-persistence theorem states that if the optimal single-copy estimation cost lies strictly above the Holevo bound, the optimal cost remains above the bound for any finite number of copies~\cite{Conlon2022gap}. Whether this persistence extends to three or more parameters has remained an open question~\cite{Conlon2022gap}. A counterexample would establish that an asymptotic precision limit can become exactly attainable with finite quantum resources. It would also raise the questions of which quantum models permit this behavior and whether the associated precision advantage survives experimental noise.

Here we address this question by constructing an explicit three-parameter estimation model and  experimentally implementing its collective measurement in a superconducting-qubit system. The model has rank two in a four-dimensional Hilbert space, and its Holevo bound is unattainable with one copy yet attained exactly by a collective measurement on two copies. The gap therefore closes without collective measurements on larger ensembles, establishing that finite-copy gap persistence is not universal beyond two parameters. The model also closes a strict gap between the Nagaoka--Hayashi and Holevo bounds, resolving the higher-parameter persistence question for these bounds~\cite{Conlon2021,Conlon2022gap}. We further prove that the gap between the attainable estimation cost and the Holevo bound persists for full-rank models with any number of parameters and any finite number of copies, identifying rank deficiency as a necessary condition for this gap to close. 
Experimentally, we observe a two-copy estimation error below single-copy and separate-measurement lower bounds that are evaluated using state tomography. Controlled depolarization removes exact finite-copy attainability while preserving a two-copy advantage over a finite noise range. 
Together, these results show that a two-copy measurement can attain the Holevo bound and retain an advantage over separate measurements even when noise prevents exact attainment.

\textit{Framework}. We first present the local estimation settings and introduce relevant quantum estimation bounds. Let $\rho_{\bm{\theta}}$ be a smooth, locally identifiable $p$-parameter model on a finite-dimensional Hilbert space $\mathcal H$, with $\bm{\theta}=(\theta_1,\ldots,\theta_p)^T$ and $\bm{\theta}=0$ an interior point. We write $\rho=\rho_{\bm 0}$ and tangent operators $D_\mu=\left.\partial_{\theta_\mu}\rho_{\bm{\theta}}\right|_{\bm{\theta}=0}$. A positive operator-valued measurement $\mathcal{M}=\{M_k\}$ on $n$ copies assigns an estimator $\widetilde{\bm{\theta}}(k)$ to outcome $k$. Its mean-square error matrix is $V$, and the estimator is locally unbiased when its mean vanishes and its first derivative is the identity at the local point. For a real symmetric positive-definite weight matrix $W=W^T>0$, the normalized finite-copy optimum, which gives the smallest normalized weighted estimation error attainable with measurements on $n$ copies, is defined as
\begin{equation}
 \cn{n}(W)=n\min_{\mathcal{M},\widetilde{\bm{\theta}}}\Tr(WV).
 \label{eq:Cn}
\end{equation}
Here the minimization is over locally unbiased estimators for the local model $(\rho,D_1,\ldots,D_p)$. The quantity $\cn{n}$ is also known as the normalized most informative bound for measurements on $n$ copies~\cite{Holevo2011,HayashiOuyang2023,Conlon2022gap}. 

To bound $\cn{n}$ from below, we introduce Hermitian estimator operators $X=(X_1,\ldots,X_p)$ satisfying $\Tr(\rho X_\mu)=0$ and $\Tr(D_\nu X_\mu)=\delta_{\mu\nu}$, and define $Z_{\mu\nu}(X)=\Tr(\rho X_\mu X_\nu)$. The Holevo Cram\'{e}r-Rao bound is then given by~\cite{Holevo2011,Demkowicz2020}
\begin{equation}
 \begin{aligned}
 \cH(W)=\min_X\big\{&\Tr[W\,\operatorname{Re}Z(X)]\\
 &+\norm{\sqrt W\,\operatorname{Im}Z(X)\sqrt W}_1\big\},
 \end{aligned}
 \label{eq:Holevo}
\end{equation}
where the trace norm is defined as $\norm{A}_1=\Tr\sqrt{A^\dagger A}$.
The Holevo bound sets the asymptotically attainable limit on the normalized weighted mean-square error when collective measurements on many identical copies are allowed~\cite{Yamagata2013,Yang2019}.
It obeys $\cn{n}(W)\geq \cH(W)$, while the Holevo bound of the $n$-copy model equals $\cH(W)/n$~\cite{Yamagata2013,Yang2019,Conlon2022gap}.
For finite-dimensional models, Eq.~\eqref{eq:Holevo} can be evaluated by semidefinite programming~\cite{Albarelli2019}.

We also write $C_{\mathrm{NH}}^{(n)}(W)$ for $n$ times the Nagaoka--Hayashi (NH) bound evaluated on $\rho_{\bm{\theta}}^{\otimes n}$, with the $n$ copies treated as one probe~\cite{Nagaoka2005,Hayashi1999,Conlon2021,HayashiOuyang2023}. 
The NH bound gives a computable lower bound on the normalized estimation error achievable with measurements on an $n$-copy block, but need not be attainable~\cite{Conlon2025extended}.
With the above normalization, these bounds satisfy~\cite{Conlon2021,HayashiOuyang2023}
\begin{equation}
 \cH(W)\leq C_{\mathrm{NH}}^{(n)}(W)\leq\cn{n}(W).
 \label{eq:NHhierarchy}
\end{equation}
The operator-moment constraints defining this semidefinite relaxation are given in the Supplemental Material~\cite{SupplementaryMaterial}.
The finite-copy persistence question is whether a strict single-copy gap, $\cn{1}>\cH$, necessarily implies $\cn{n}>\cH$ for each finite $n$ (we omit \(W\) from the notation when the model and weight are fixed hereafter). We show that this implication fails with the following model.

\textit{Three-parameter counterexample}. We write $\id$ for the identity operator, with its space indicated by a subscript. We consider the four-dimensional Hilbert space $\mathcal{H}=\mathbb{C}_r^2\otimes\mathbb{C}_t^2$, where $r$ and $t$ label the two qubits, together with the weight matrix $W=\id_3$. Let
\begin{equation}
\rho_0=|0\rangle\!\langle0|_r\otimes\frac{\id_t}{2},\qquad
\rho_{\bm{\theta}}=U_{\bm{\theta}}\rho_0U_{\bm{\theta}}^\dagger,
 \label{eq:model}
\end{equation}
where
\begin{equation}
 \begin{aligned}
 U_{\bm{\theta}}=\exp\!\big[-\ii\sqrt{2}\big(&
 \theta_1\sigma_x^{(r)}\otimes\id_t+
 \theta_2\sigma_y^{(r)}\otimes\id_t\\
&+\theta_3\sigma_y^{(r)}\otimes\sigma_x^{(t)}\big)\big],
 \end{aligned}
 \label{eq:unitary}
\end{equation}
encodes the parameter $\bm{\theta}=(\theta_1,\theta_2,\theta_3)^T$ of interest and the Pauli operators act on the indicated tensor factors. The resulting state family is smooth, positive, and has constant rank two. Its tangent operators at $\bm{\theta}=0$ are linearly independent, so the model is locally identifiable.

To evaluate the Holevo bound, we decompose the estimator operators into blocks on the support of $\rho_0$ and its kernel. A trace-norm inequality bounds the Holevo functional from below by a quadratic form in these blocks. Local unbiasedness and the Cauchy--Schwarz inequality then bound this quadratic form below by $5/8$ for all feasible estimator operators. The Supplemental Material~\cite{SupplementaryMaterial} gives the derivation and an explicit set of locally unbiased estimator operators whose Holevo functional equals $5/8$. The lower bound and explicit construction together establish
\begin{equation}
 \cH(\id_3)=\frac58.
 \label{eq:HolevoValue}
\end{equation}
We also characterize the complete family of estimator operators that attain this value in the Supplemental Material~\cite{SupplementaryMaterial}.

We next determine the exact single-copy optimum. The state is block diagonal in the $\sigma_x^{(t)}$ basis, so we can first measure $t$ in this basis without losing information. Each outcome $a=\pm1$ leaves a pure-qubit model on $r$ with coordinates $(\theta_1,\phi_a)$, where $\phi_a=\theta_2+a\theta_3$. We therefore only need to optimize the conditional qubit measurements. Each conditional model has quantum Fisher information $8\id_2$, so the Gill--Massar inequality bounds the trace of its classical Fisher information matrix by $8$~\cite{GillMassar2000}. We use symmetry and convexity to choose identical conditional Fisher information matrices for $a=+1$ and $a=-1$ without increasing the classical Cram\'{e}r-Rao cost. Minimizing the three-parameter Cram\'{e}r-Rao bound under this information constraint then gives a lower bound on $\cn{1}$. We attain this bound by measuring $\sigma_y^{(r)}$ with probability $1/(1+\sqrt2)$ and $\sigma_x^{(r)}$ otherwise, retaining the outcome $a$ and using the local score estimator. This establishes
\begin{equation}
 \cn{1}(\id_3)=\frac{3+2\sqrt2}{8}>\frac58.
 \label{eq:singleCopyValue}
\end{equation}
The above result shows the gap between the attainable single-copy optimum and the Holevo bound for the studied three-parameter model.
In the Supplemental Material~\cite{SupplementaryMaterial}, we also give an independent proof using a Naimark dilation~\cite{Holevo2011}, which represents any measurement by commuting observables on an enlarged Hilbert space and shows that this commutativity is incompatible with the Holevo equality conditions for one copy.

\begin{figure*}[t]
\includegraphics[width=\textwidth]{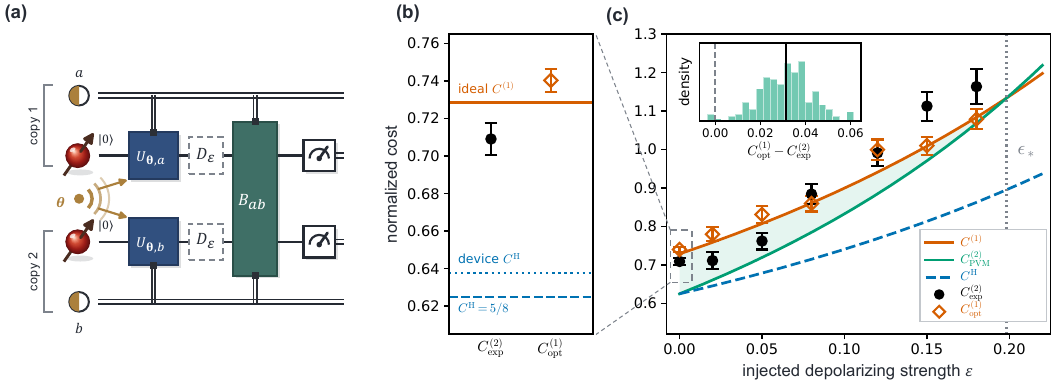}
\caption{Experimental demonstration on \texttt{ibm\_boston}. \textbf{(a)}~Hybrid two-transmon protocol. The classical flags $a$ and $b$ are sampled uniformly and select the conditional encodings $U_{\bm{\theta},a}$ and $U_{\bm{\theta},b}$, the optional dashed gates inject the depolarizing strength $\epsilon$, and the sector-selected basis change $B_{ab}$ followed by computational-basis readout implements the collective measurement of Eqs.~\eqref{eq:rbasis} and \eqref{eq:twoCopyX}. 
\textbf{(b)}~At $\epsilon=0$, the measured cost $C^{(2)}_{\mathrm{exp}}$ (filled circle) lies below the certified single-copy lower bound $C^{(1)}_{\mathrm{opt}}$ (open diamond), with the ideal $\cn{1}$ (solid orange), the reconstructed device Holevo bound of $0.638$ (dotted blue), and $\cH=5/8$ (dashed blue) as references. 
\textbf{(c)}~Measured costs against the injected depolarizing strength $\epsilon$. Curves are the exact single-copy optimum $\cn{1}(\epsilon)$, the cost $\cn{2}_{\mathrm{PVM}}(\epsilon)$ of the ideal two-copy projection-valued measurement, and the Holevo bound $\cH(\epsilon)$ of the depolarized model. The shaded region marks the ideal advantage window, and the dotted line marks its crossover $\epsilon_*$. Inset:~Bootstrap distribution of the advantage $C^{(1)}_{\mathrm{opt}}-C^{(2)}_{\mathrm{exp}}$ at $\epsilon=0$, with zero (dashed) and the point estimate (solid) marked. Error bars throughout are bootstrap standard errors, and the shot counts for each data point are given in~\cite{SupplementaryMaterial}.}
\label{fig:experiment}
\end{figure*}

We now demonstrate that two copies remove this obstruction. On two copies, we label the qubits of each copy by $(r_1,t_1)$ and $(r_2,t_2)$. Measuring $t_1$ and $t_2$ in the $\sigma_x$ basis gives outcomes $a,b\in\{+1,-1\}$. For the $r_1r_2$ factors, we use the ordered basis
\begin{equation}
 \left(|00\rangle,|u_+\rangle,|u_-\rangle,|11\rangle\right),\qquad
 |u_\pm\rangle=\frac{|10\rangle\pm|01\rangle}{\sqrt2}.
 \label{eq:rbasis}
\end{equation}
In this order, the following blockwise estimator observables act as Pauli operators on two effective qubits:
\begin{equation}
\begin{aligned}
ab=+1:\;&X_1^{(2)}=0,\;
X_2^{(2)}=\frac{\id\otimes\sigma_x}{2},\;
X_3^{(2)}=\frac{a\,\id\otimes\sigma_x}{4},\\
ab=-1:\;&X_1^{(2)}=-\frac{\id\otimes\sigma_y}{2},\;
X_2^{(2)}=0,\;
X_3^{(2)}=\frac{a\,\sigma_x\otimes\id}{4}.
\end{aligned}
\label{eq:twoCopyX}
\end{equation}
The two nonzero observables in the $ab=-1$ block act on different effective qubits. All three observables therefore commute, and their common spectral measure is a projective measurement. Writing $D_\mu^{(2)}=\left.\partial_{\theta_\mu}\rho_{\bm{\theta}}^{\otimes2}\right|_{\bm{\theta}=0}$, direct evaluation gives~\cite{SupplementaryMaterial}
\begin{equation}
 \begin{aligned}
 \Tr(\rho_0^{\otimes2}X_\mu^{(2)})&=0,\qquad
 \Tr(D_\mu^{(2)}X_\nu^{(2)})=\delta_{\mu\nu},\\
 V^{(2)}&=\operatorname{diag}\!\left(\frac18,\frac18,\frac1{16}\right).
 \end{aligned}
 \label{eq:twoCopyMoments}
\end{equation}
Here $V^{(2)}$ is the mean-square error matrix of the two-copy estimator. The measurement is therefore locally unbiased, and its normalized cost equals $2\Tr V^{(2)}=5/8$. Combining this result with Eqs.~\eqref{eq:HolevoValue} and \eqref{eq:singleCopyValue} gives
\begin{equation}
 \cn{1}=\frac{3+2\sqrt2}{8}>\cH=\cn{2}=\frac58.
 \label{eq:counterexample}
\end{equation}
Equation~\eqref{eq:counterexample} therefore disproves universal finite-copy gap persistence between the most informative and Holevo bounds using the studied three-parameter model.
We remark that three parameters and two copies are the minimum numbers needed for closing a strict single-copy gap.
We also list the joint projectors and their estimator values explicitly in the Supplemental Material~\cite{SupplementaryMaterial}.

For this model, the single-copy Nagaoka--Hayashi bound is attainable and therefore equals the exact single-copy optimum.
In~\cite{SupplementaryMaterial}, we derive upper and lower bounds on $C_{\mathrm{NH}}^{(1)}$ that both equal $\cn{1}$, establishing $C_{\mathrm{NH}}^{(1)}=\cn{1}$.
Combining this equality with Eq.~\eqref{eq:NHhierarchy} and the two-copy measurement gives
\begin{equation}
 \begin{aligned}
 C_{\mathrm{NH}}^{(1)}&=\cn{1}=\frac{3+2\sqrt2}{8},\\
 C_{\mathrm{NH}}^{(2)}&=\cn{2}=\cH=\frac58.
 \end{aligned}
 \label{eq:NHclosure}
\end{equation}
The strict gap between the Nagaoka--Hayashi and Holevo bounds therefore also closes at two copies, disproving weak gap persistence for these bounds beyond two parameters~\cite{Conlon2022gap}.

\textit{Full-rank models.} We now establish that rank deficiency is essential for the above phenomenon, and show that in any finite-dimensional full-rank model, a strict single-copy gap between the most-informative and Holevo bounds persists at any finite copy number. Therefore, full rank rules out the finite-copy gap closure exhibited above.
\begin{theorem}
\label{thm:fullrank}
Let $\rho_{\bm{\theta}}$ be a smooth, locally identifiable $p$-parameter model on a finite-dimensional Hilbert space, with $\bm{\theta}=0$ an interior point and $\rho=\rho_{\bm 0}>0$ a full-rank state. For any real symmetric weight matrix $W=W^T>0$ and each integer $n\geq1$,
\begin{equation}
 \begin{aligned}
 \cn{1}(W)&>\cH(W) 
 \Longrightarrow\
 \cn{n}(W)>\cH(W).
 \end{aligned}
 \label{eq:fullrank}
\end{equation}
\end{theorem}

We provide the detailed proof in the Supplemental Material~\cite{SupplementaryMaterial} and outline the main argument here. We first show that the finite-copy optimization has an optimal measurement. Assume, for contradiction, that this measurement attains the Holevo bound for some finite $n$. We use a Naimark dilation to represent it by commuting estimator observables on an enlarged Hilbert space~\cite{Holevo2011}. Because the state has full rank, equality with the Holevo bound eliminates the additional noise introduced by the dilation. The corresponding estimator operators on the original $n$-copy space must therefore commute. Additivity and strict convexity make the averaged one-copy optimizer the unique $n$-copy optimizer, so its commutativity implies, after tracing out $n-1$ copies, that the one-copy optimizer also commutes. We could therefore measure its operators jointly and attain the Holevo bound with one copy, contradicting $\cn{1}>\cH$.

\textit{Experimental demonstration}.
We experimentally realized the three-parameter model and its two-copy measurement using transmons of the IBM superconducting processor \texttt{ibm\_boston}. Since $[U_{\bm{\theta}},\sigma_x^{(t)}]=0$, the flag $a$ introduced above can be sampled classically and the conditional rotation applied to a single transmon, which reproduces the single-copy outcome statistics exactly.  As shown in Fig.~\ref{fig:experiment}(a), two copies then require two transmons and two recorded bits $(a,b)$, and for each label pair the measurement of Eqs.~\eqref{eq:rbasis} and \eqref{eq:twoCopyX} compiles to single-qubit rotations and two CZ gates. 

The normalized two-copy mean-square-error cost $C^{(2)}_{\mathrm{exp}}$ is estimated in two stages.
We first calibrate the estimator from its response to small positive and negative parameter shifts and independently measure its covariance at the local point $\bm{\theta}=0$. We then use the measured response and covariance to estimate $C^{(2)}_{\mathrm{exp}}$.
In parallel, we use state tomography to reconstruct the prepared states and their response to small parameter changes. Applying the Gill--Massar bound to the reconstructed model gives the single-copy lower bound $C^{(1)}_{\mathrm{opt}}$. The gate decompositions, reconstruction procedure and assumptions, and statistical analysis are described in the Supplemental Material~\cite{SupplementaryMaterial}.

We now show the comparison between the measured two-copy cost $C^{(2)}_{\mathrm{exp}}$ and a certified single-copy cost lower bound $C^{(1)}_{\mathrm{opt}}$. From $4.1\times10^6$ protocol shots and $7.3\times10^6$ tomography shots~\cite{SupplementaryMaterial}, we obtain $C^{(2)}_{\mathrm{exp}}=0.709(9)$ against $C^{(1)}_{\mathrm{opt}}=0.741(7)$, a two-copy advantage of $0.032(11)$, or $2.8$ standard deviations, as shown in Fig.~\ref{fig:experiment}(b). 
The inset in Fig.~\ref{fig:experiment}(c) shows the bootstrap distribution of $C^{(1)}_{\mathrm{opt}}-C^{(2)}_{\mathrm{exp}}$, whose displacement above zero quantifies the statistical evidence for a collective advantage.
The measured two-copy cost $C_{\mathrm{exp}}^{(2)}$ also lies $2.3$ standard deviations below the exact single-copy optimum of the ideal model, $\cn{1}=(3+2\sqrt{2})/8\simeq0.729$, which  carries no reconstruction uncertainty.
Under the calibrated readout model, the outcome distribution inferred after readout-error mitigation gives a two-copy cost of $0.698(8)$~\cite{Bravyi2021}, as detailed in the Supplemental Material~\cite{SupplementaryMaterial}.

The comparisons above show a cost below the single-copy limits of the reconstructed models. To determine whether this advantage genuinely requires a collective measurement, we must also rule out strategies that measure the two copies separately and combine their outcomes. Because hardware imperfections affect the two transmons differently, we reconstruct a slightly different local model for each. For this pair of reconstructed models, we derive a lower bound for all such separate-measurement strategies using the Gill--Massar constraints~\cite{SupplementaryMaterial}.
The resulting separate-measurement bound is $0.778(8)$. The measured collective cost $C_{\mathrm{exp}}^{(2)}$  lies below it by $5.9$ standard deviations. This comparison demonstrates a two-copy collective advantage, while the measured cost remains above the ideal Holevo value $\cH=5/8$. 
From state tomography, we further obtain a Holevo bound of $0.638$ for the reconstructed single-copy model used to evaluate $C^{(1)}_{\mathrm{opt}}$.

To determine how far the advantage survives noise, we replace each prepared copy by $(1-\epsilon)\rho_{\bm{\theta}}+\epsilon\,\id/4$ using randomized Pauli gates drawn independently on the two copies~\cite{SupplementaryMaterial}. The depolarized model leads to closed-form bounds, $\cn{1}(\epsilon)=(3+2\sqrt2)/[8(1-\epsilon)^2]$ and $\cH(\epsilon)=(5-2\epsilon)/[8(1-\epsilon)^2]$, and the cost of the fixed measurement of Eq.~\eqref{eq:twoCopyX} with its estimator reoptimized at each $\epsilon$ crosses $\cn{1}(\epsilon)$ at $\epsilon_*\simeq0.198$, closing the ideal advantage window. For any $\epsilon>0$ the prepared state is full rank, so Theorem~\ref{thm:fullrank} excludes exact attainment of the Holevo bound at any copy number. Thus, arbitrarily weak depolarization rules out exact finite-copy saturation, while a two-copy precision advantage survives over a finite noise range. Fig.~\ref{fig:experiment}(c) shows the measured costs across the $\epsilon$-sweep. The measured cost lies below the reconstructed single-copy lower bound by $2.2$ standard deviations at $\epsilon=0.02$ and $0.05$. At the larger sampled noise strengths, it is no longer significantly below this bound. 
At $\epsilon=0$, the excess of the measured two-copy cost over the ideal cost of the two-copy projective measurement $\cn{2}_{\mathrm{PVM}}(0)$ is comparable to that produced by an injected depolarizing strength of approximately $0.05$, consistent with the intrinsic device noise.

\textit{Discussion}. For our three-parameter model, both the most informative and Nagaoka--Hayashi bounds have a strict single-copy gap above the Holevo bound. Their exact closure at two copies establishes that finite-copy gap persistence is not universal in multiparameter quantum estimation~\cite{Conlon2022gap}. The most informative and Nagaoka--Hayashi bounds coincide at both one and two copies, so this example does not resolve persistence of a gap between them. The counterexample shows that a strict single-copy gap can close with two copies, while Theorem~\ref{thm:fullrank} rules out such closure for full-rank models with any number of parameters. Together, these results identify rank deficiency as a necessary condition for finite-copy gap closure.

We now provide the physical intuition for the presented counterexample, whose mechanism is tied to the first-order motion of the state support. At the local point, parameter changes connect the rank-two support to its two-dimensional kernel, creating estimator components that are absent in a full-rank model. The third parameter labels complementary internal sectors, allowing two copies to distribute commutation constraints that conflict on one copy. When the support is stationary to first order, the local model reduces to a full-rank model on its support and Theorem~\ref{thm:fullrank} applies. Rank deficiency and support motion are thus necessary, but we have not shown them to be sufficient for finite-copy closure. This distinction also clarifies numerical near-saturation in noisy three-parameter magnetometry~\cite{Friel2020}. 
When dephasing makes the state full rank and a strict single-copy gap remains, Theorem~\ref{thm:fullrank} guarantees a positive gap at any finite copy number, even when numerical results suggest near-saturation.

The superconducting-qubit experiment in this work provides an operational interpretation of the two-copy construction. 
Our hybrid implementation reproduces the target outcome statistics exactly while representing the $t$ subsystems by sampled classical flags. 
We observed a two-copy estimation error below the separate-measurement lower bound evaluated for the reconstructed models, providing evidence for a collective advantage under the reconstruction assumptions. We note that exact Holevo bound saturation occurs only in the noiseless model, while a collective precision advantage remains observable over a finite noise range. 
Beyond the explicit construction in this work, it remains open which rank-deficient models admit finite-copy closure and how to realize this property in physically motivated sensing models.
Our results clarify how collective measurements on finite ensembles can access the ultimate precision limits of quantum multiparameter estimation.

\textit{Acknowledgments}. We thank Shihao Ru, Paul Nation and the Unitary Foundation team for providing helpful feedback on the manuscript. We thank IBM for providing quantum cloud computing credits.

\bibliography{main}

\clearpage
\onecolumngrid
\clearpage



\section*{Supplementary material}

# Supplemental Material for "Closing a single-copy gap to the Holevo bound with two-copy measurements"

Changhao Li[1,*]

[1] *Unitary Foundation, San Francisco, CA, USA*



## CONTENTS

# I. LOCAL ESTIMATION FRAMEWORK

We first present the details of the local estimation framework and useful definitions. Let $\rho_{\boldsymbol{\theta}}$ be a smooth, locally identifiable $p$-parameter family of density operators on a finite-dimensional Hilbert space $\mathcal{H}$, with $\boldsymbol{\theta} = 0$ an interior point. For parameter $\mu$, we write

$$\rho = \rho_{\mathbf{0}}, \qquad D_\mu = \partial_{\theta_\mu} \rho_{\boldsymbol{\theta}} \big|_{\boldsymbol{\theta}=\mathbf{0}} .$$

The tangent operators $D_1, \ldots, D_p$ are therefore linearly independent. We write $\mathbb{I}$ for the identity operator, with its space indicated by a subscript or clear from the tensor context. For $n$ copies, the state and tangent operators are

$$\rho^{(n)} = \rho^{\otimes n}, \qquad D_\mu^{(n)} = \sum_{m=1}^{n} \rho^{\otimes (m-1)} \otimes D_\mu \otimes \rho^{\otimes (n-m)}.$$

A positive operator-valued measurement (POVM) $\mathcal{M} = \{M_k\}$ assigns a real estimator $\widetilde{\boldsymbol{\theta}}(k)$ to outcome $k$. After translating the local point to zero, local unbiasedness is

$$\sum_k \mathrm{Tr}(\rho^{(n)} M_k) \widetilde{\theta}_\mu(k) = 0, \qquad \sum_k \mathrm{Tr}(D_\nu^{(n)} M_k) \widetilde{\theta}_\mu(k) = \delta_{\mu\nu}.$$

The mean-square error matrix is

$$V_{\mu\nu} = \sum_k \mathrm{Tr}(\rho^{(n)} M_k) \widetilde{\theta}_\mu(k) \widetilde{\theta}_\nu(k).$$

All estimation costs and bounds below are evaluated at $\boldsymbol{\theta} = 0$, with this dependence suppressed in the notations. For a real symmetric positive-definite weight matrix $W = W^T > 0$, the normalized most informative (MI) bound for measurements on $n$ copies is [1–3]

$$C^{(n)}(W) = n \min_{\mathcal{M}, \widetilde{\boldsymbol{\theta}}} \mathrm{Tr}(WV). \tag{S1}$$

For full-rank finite-dimensional models, the minimum in Eq. (S1) is attained. This attainment statement is proved at the beginning of Sec. III and is used only there.

The Holevo optimization uses Hermitian estimator operators $X = (X_1, \ldots, X_p)$ satisfying

$$\mathrm{Tr}(\rho X_\mu) = 0, \qquad \mathrm{Tr}(D_\nu X_\mu) = \delta_{\mu\nu}. \tag{S2}$$

With $Z_{\mu\nu}(X) = \mathrm{Tr}(\rho X_\mu X_\nu)$ and the trace norm $\|A\|_1 = \mathrm{Tr}\sqrt{A^\dagger A}$, the Holevo bound is [1, 4]

$$C^{\mathrm{H}}(W) = \min_X \left\{ \mathrm{Tr}[W \,\mathrm{Re}\, Z(X)] + \left\| \sqrt{W} \,\mathrm{Im}\, Z(X) \sqrt{W} \right\|_1 \right\}. \tag{S3}$$

We omit the weight argument in the notations once the local model and $W$ have been fixed. For the $n$-copy model, the Holevo bound equals $C^{\mathrm{H}}(W)/n$ and is asymptotically attainable under the standard regularity conditions [4–6]. We use the corresponding normalization for the Nagaoka–Hayashi (NH) bound, denoted $C_{\mathrm{NH}}^{(n)}(W)$, so that

$$C^{\mathrm{H}}(W) \le C_{\mathrm{NH}}^{(n)}(W) \le C^{(n)}(W). \tag{S4}$$

Here the $n$ copies are treated as one probe when evaluating the Nagaoka–Hayashi bound [2, 7]. Its semidefinite formulation and exact value for the presented counterexample in this work are given in Sec. II C.

Any measurement and estimator admit a Naimark representation that will be useful throughout [1]. Let $\mathcal{V} : \mathcal{H} \to \widetilde{\mathcal{H}}$ be an isometry such that $M_k = \mathcal{V}^\dagger \widetilde{M}_k \mathcal{V}$ for orthogonal projectors $\widetilde{M}_k$. The estimator observables

$$\widetilde{X}_\mu = \sum_k \widetilde{\theta}_\mu(k) \widetilde{M}_k$$

are Hermitian and pairwise commuting. Their compressed operators $X_\mu = \mathcal{V}^\dagger \widetilde{X}_\mu \mathcal{V}$ satisfy Eq. (S2). The covariance of the measurement is the second moment of the commuting observables $\widetilde{X}_\mu$.

## II. THREE-PARAMETER COUNTEREXAMPLE

### A. Model and Holevo bound

We provide the details of the three-parameter model introduced in the main text and derive its Holevo bound. We write the single-copy Hilbert space as $\mathcal{H} = \mathbb{C}_r^2 \otimes \mathbb{C}_t^2$, where we use $r$ and $t$ to label the two qubits. The initial probe state and the three Hermitian parameter generators are

$$\rho_0 = |0\rangle\langle 0|_r \otimes \frac{\mathbb{I}_t}{2}, \quad \mathcal{G}_1 = \sqrt{2}\,\sigma_x^{(r)} \otimes \mathbb{I}_t, \quad \mathcal{G}_2 = \sqrt{2}\,\sigma_y^{(r)} \otimes \mathbb{I}_t, \quad \mathcal{G}_3 = \sqrt{2}\,\sigma_y^{(r)} \otimes \sigma_x^{(t)}.$$

We define the model as

$$\rho_{\boldsymbol{\theta}} = U_{\boldsymbol{\theta}} \rho_0 U_{\boldsymbol{\theta}}^\dagger, \qquad U_{\boldsymbol{\theta}} = \exp\left[-\mathrm{i}\sum_{\mu=1}^{3} \theta_\mu \mathcal{G}_\mu\right]. \tag{S5}$$

The resulting state family is smooth, positive for all real $\boldsymbol{\theta}$, and has constant rank two. Direct differentiation gives the tangent operators that are linearly independent:

$$D_1 = -\frac{\sigma_y^{(r)} \otimes \mathbb{I}_t}{\sqrt{2}}, \qquad D_2 = \frac{\sigma_x^{(r)} \otimes \mathbb{I}_t}{\sqrt{2}}, \qquad D_3 = \frac{\sigma_x^{(r)} \otimes \sigma_x^{(t)}}{\sqrt{2}}. \tag{S6}$$

We let $P = |0\rangle\langle 0|_r \otimes \mathbb{I}_t$ and $Q = \mathbb{I} - P$. The support $P\mathcal{H}$ consists of vectors $|0\rangle_r \otimes |\psi\rangle_t$, while the kernel $Q\mathcal{H}$ consists of vectors $|1\rangle_r \otimes |\psi\rangle_t$, with $|\psi\rangle_t \in \mathbb{C}_t^2$. Using the $t$-qubit basis in each subspace, we represent the maps $QD_\mu P$ from the support to the kernel as $2 \times 2$ matrices:

$$\Gamma = QD_2P = \frac{\mathbb{I}_t}{\sqrt{2}}, \qquad \Gamma_x = QD_3P = \frac{\sigma_x^{(t)}}{\sqrt{2}}, \qquad QD_1P = -\mathrm{i}\Gamma. \tag{S7}$$

With the Hilbert–Schmidt inner product $\langle A, B\rangle_{\mathrm{HS}} = \mathrm{Tr}(A^\dagger B)$ and normalization $\|A\|_{\mathrm{HS}} = \sqrt{\langle A, A\rangle_{\mathrm{HS}}}$, they obey

$$\|\Gamma\|_{\mathrm{HS}} = \|\Gamma_x\|_{\mathrm{HS}} = 1, \qquad \langle \Gamma, \Gamma_x\rangle_{\mathrm{HS}} = 0. \tag{S8}$$

We consider the weight matrix to be $W = \mathbb{I}_3$. For locally unbiased estimator operators $X_\mu$, define the support and support-to-kernel blocks to be $\mathcal{A}_\mu = PX_\mu P$ and $\mathcal{B}_\mu = QX_\mu P$, respectively. The kernel-to-kernel block does not enter either the objective or the local unbiasedness conditions. Consider the quadratic functional

$$q(X) = \mathrm{Tr}[\rho_0(X_1 + \mathrm{i}X_2)(X_1 - \mathrm{i}X_2)] + \mathrm{Tr}(\rho_0 X_3^2).$$

With $Z_{\mu\nu}(X) = \mathrm{Tr}(\rho X_\mu X_\nu)$, this functional is

$$q(X) = \mathrm{Tr}\,\mathrm{Re}\,Z(X) + 2[\mathrm{Im}\,Z(X)]_{12}.$$

Since $\mathrm{Im}\,Z(X)$ is a real antisymmetric $3 \times 3$ matrix, its trace norm satisfies

$$\|\mathrm{Im}\,Z(X)\|_1 = 2\sqrt{\sum_{\mu<\nu} [\mathrm{Im}\,Z(X)]_{\mu\nu}^2} \geq 2[\mathrm{Im}\,Z(X)]_{12}.$$

The Holevo functional $h(X) = \mathrm{Tr}\,\mathrm{Re}\,Z(X) + \|\mathrm{Im}\,Z(X)\|_1$ therefore satisfies $h(X) \geq q(X)$.

To bound $q(X)$ using the local unbiasedness constraints, we express it in terms of the support blocks. Since $\rho_0 = P/2$, we have $\mathrm{Tr}(\rho_0 A^\dagger A) = \|AP\|_{\mathrm{HS}}^2/2$ for any operator $A$. Thus

$$q(X) = \frac{1}{2}\left(\|(X_1 - \mathrm{i}X_2)P\|_{\mathrm{HS}}^2 + \|X_3 P\|_{\mathrm{HS}}^2\right).$$

Decomposing the output of each map into its support and kernel components gives

$$q(X) = \frac{1}{2}\left(\|\mathcal{A}_1 - \mathrm{i}\mathcal{A}_2\|_{\mathrm{HS}}^2 + \|\mathcal{B}_1 - \mathrm{i}\mathcal{B}_2\|_{\mathrm{HS}}^2 + \|\mathcal{A}_3\|_{\mathrm{HS}}^2 + \|\mathcal{B}_3\|_{\mathrm{HS}}^2\right). \tag{S9}$$

The local unbiasedness conditions can be expressed as

$$-2\,\mathrm{Im}\langle\Gamma,\mathcal{B}_\nu\rangle_{\mathrm{HS}}=\delta_{1\nu},\qquad 2\,\mathrm{Re}\langle\Gamma,\mathcal{B}_\nu\rangle_{\mathrm{HS}}=\delta_{2\nu},\qquad 2\,\mathrm{Re}\langle\Gamma_x,\mathcal{B}_\nu\rangle_{\mathrm{HS}}=\delta_{3\nu}. \tag{S10}$$

In particular, $\langle\Gamma,\mathcal{B}_1-\mathrm{i}\mathcal{B}_2\rangle_{\mathrm{HS}}=-\mathrm{i}$ and $\mathrm{Re}\langle\Gamma_x,\mathcal{B}_3\rangle_{\mathrm{HS}}=\frac{1}{2}$.

Cauchy-Schwarz and Eq. (S8) imply

$$\|\mathcal{B}_1-\mathrm{i}\mathcal{B}_2\|_{\mathrm{HS}}\geq 1,\qquad \|\mathcal{B}_3\|_{\mathrm{HS}}\geq\frac{1}{2}. \tag{S11}$$

Combining Eqs. (S9) and (S11) gives

$$h(X)\geq q(X)\geq\frac{1}{2}\left(1+\frac{1}{4}\right)=\frac{5}{8}. \tag{S12}$$

The lower bound is attained by

$$\mathcal{A}_1=\mathcal{A}_2=\mathcal{A}_3=0,\qquad \mathcal{B}_1=-\frac{\mathrm{i}}{2}\Gamma,\qquad \mathcal{B}_2=\frac{1}{2}\Gamma,\qquad \mathcal{B}_3=\frac{1}{2}\Gamma_x.$$

For the Hermitian completions $PX_\mu Q=\mathcal{B}_\mu^\dagger$, the complex covariance matrix is

$$Z=\begin{pmatrix}1/8 & \mathrm{i}/8 & 0\\ -\mathrm{i}/8 & 1/8 & 0\\ 0 & 0 & 1/8\end{pmatrix}.$$

Thus $\mathrm{Tr}\,\mathrm{Re}\,Z=3/8$ and $\|\mathrm{Im}\,Z\|_1=1/4$. This gives $C^{\mathrm{H}}(\mathbb{I}_3)\leq h(X)=5/8$, which, together with the lower bound in Eq. (S12), establishes

$$C^{\mathrm{H}}(\mathbb{I}_3)=\frac{5}{8}. \tag{S13}$$

We will also need all equality conditions. Saturation of the lower-bound chain in Eq. (S12) requires $q(X)=5/8$. Equations (S9) and (S11) then force $\mathcal{A}_1-\mathrm{i}\mathcal{A}_2=0$ and $\mathcal{A}_3=0$. Since the support blocks are Hermitian, $\mathcal{A}_1=\mathcal{A}_2=\mathcal{A}_3=0$. Equality in the two Cauchy-Schwarz bounds gives

$$\mathcal{B}_1-\mathrm{i}\mathcal{B}_2=-\mathrm{i}\Gamma,\qquad \mathcal{B}_3=\frac{1}{2}\Gamma_x.$$

The general solution consistent with Eq. (S10) is

$$\mathcal{B}_1=-\frac{\mathrm{i}}{2}\Gamma+R,\qquad \mathcal{B}_2=\frac{1}{2}\Gamma-\mathrm{i}R,\qquad \mathcal{B}_3=\frac{1}{2}\Gamma_x, \tag{S14}$$

where $R:P\mathcal{H}\to Q\mathcal{H}$ is a linear map satisfying

$$\langle\Gamma,R\rangle_{\mathrm{HS}}=0,\qquad \langle\Gamma_x,R\rangle_{\mathrm{HS}}=0. \tag{S15}$$

For this family, direct evaluation gives

$$\mathrm{Tr}\,\mathrm{Re}\,Z=\frac{3}{8}+\|R\|_{\mathrm{HS}}^2,\qquad \|\mathrm{Im}\,Z\|_1=\left|\frac{1}{4}-\|R\|_{\mathrm{HS}}^2\right|.$$

Modulo arbitrary Hermitian kernel-to-kernel blocks $QX_\mu Q$, the complete Holevo equality family contains the additional condition

$$\|R\|_{\mathrm{HS}}^2\leq\frac{1}{4}. \tag{S16}$$

Equations (S14), (S15), and (S16) also apply after embedding the physical Hilbert space into a Naimark dilation space [1]. In that setting $R:P\mathcal{H}\to(\mathbb{I}-P)\widetilde{\mathcal{H}}$ may have components in the added Naimark directions. We use these extended-space equality conditions in Sec. II E to rule out Holevo saturation by any single-copy POVM, using its representation by commuting observables.

### B. Exact single-copy optimum

The model has a classical-quantum decomposition that allows an exact evaluation of $C^{(1)}$. Since $U_{\boldsymbol{\theta}}$ in Eq. (S5) commutes with $\sigma_x^{(t)}$, let $|a_x\rangle$ denote the $\sigma_x^{(t)}$ eigenstate with $a \in \{+1, -1\}$. Equation (S5) becomes

$$\rho_{\boldsymbol{\theta}} = \frac{1}{2} \sum_{a=\pm 1} \rho^{(r)}_{\boldsymbol{\theta},a} \otimes |a_x\rangle\langle a_x|, \tag{S17}$$

where

$$\rho^{(r)}_{\boldsymbol{\theta},a} = U_{\boldsymbol{\theta},a}|0\rangle\langle 0|U^{\dagger}_{\boldsymbol{\theta},a}, \qquad U_{\boldsymbol{\theta},a} = \exp\left[-\mathrm{i}\sqrt{2}\left(\theta_1\sigma_x + (\theta_2 + a\theta_3)\sigma_y\right)\right]. \tag{S18}$$

Dephasing an arbitrary measurement in the $\sigma_x^{(t)}$ basis leaves all outcome probabilities unchanged. Refining its outcomes to include $a$ can only increase the classical Fisher information. Outcomes with zero probability at the local point also have zero first derivative and can be discarded. An optimal measurement may therefore be taken to measure $a$ first and then perform a conditional qubit measurement on $r$.

For fixed $a$, define $\phi_a = \theta_2 + a\theta_3$. In the local coordinates $(\theta_1, \phi_a)$, the conditional model is a pure qubit model with quantum Fisher information matrix $J_a = 8\mathbb{I}_2$. Let $I_a$ be the classical Fisher information matrix of a conditional qubit measurement. The Gill–Massar inequality gives [8]

$$\mathrm{Tr}(J_a^{-1} I_a) \leq 1, \qquad \mathrm{Tr}\, I_a \leq 8. \tag{S19}$$

The Fisher information for $\boldsymbol{\theta}$ is

$$I = \frac{1}{2} \sum_{a=\pm 1} T_a^T I_a T_a, \qquad T_a = \begin{pmatrix} 1 & 0 & 0 \\ 0 & 1 & a \end{pmatrix}. \tag{S20}$$

Swapping the conditional measurements for $a = +1$ and $a = -1$ changes the sign of the third parameter coordinate. We choose with equal probability between the original strategy and its swapped version, recording this choice as part of the measurement outcome. The resulting Fisher information is the average of the two Fisher information matrices, so the entries $I_{13}$ and $I_{23}$ cancel. Since the swapped strategy has the same cost as the original strategy, convexity of $\mathrm{Tr}\, I^{-1}$ ensures that this randomization cannot increase the cost. An optimal strategy can therefore be chosen with identical conditional Fisher information matrices in the two sectors,

$$I_{a=+1} = I_{a=-1} = \begin{pmatrix} x & c \\ c & y \end{pmatrix}, \qquad x + y \leq 8.$$

Equation (S20) then gives $I = \left(\begin{smallmatrix} x & c & 0 \\ c & y & 0 \\ 0 & 0 & y \end{smallmatrix}\right)$. For a fixed measurement, the classical Cramér-Rao bound is attained locally by the score estimator. Therefore

$$\mathrm{Tr}\, V \geq \mathrm{Tr}\, I^{-1} = \frac{x+y}{xy - c^2} + \frac{1}{y} \geq \frac{1}{x} + \frac{2}{y} \geq \frac{3 + 2\sqrt{2}}{8}.$$

We can also show that the lower bound is attainable. After measuring $a$, choose a $\sigma_y^{(r)}$ measurement with probability $1/(1+\sqrt{2})$ and a $\sigma_x^{(r)}$ measurement with probability $\sqrt{2}/(1+\sqrt{2})$. These measurements attain the conditional quantum Fisher information for $\theta_1$ and $\phi_a$, respectively. They give

$$I_a = \frac{8}{1+\sqrt{2}}\,\mathrm{diag}(1, \sqrt{2}), \qquad x = \frac{8}{1+\sqrt{2}}, \qquad y = \sqrt{2}x.$$

The local score estimator attains $V = I^{-1}$. We conclude that

$$C^{(1)}(\mathbb{I}_3) = \frac{3 + 2\sqrt{2}}{8}. \tag{S21}$$

In particular, the exact gap above the Holevo bound in Eq. (S13) equals $(\sqrt{2} - 1)/4$.

### C. Nagaoka–Hayashi bound

The Nagaoka–Hayashi (NH) bound is obtained by relaxing the operator-moment constraints of a measurement [7, 9, 10]. For the $n$-copy model, we use $X_\mu$ to denote Hermitian estimator operators on $\mathcal{H}^{\otimes n}$ satisfying the corresponding local unbiasedness conditions. Let $\mathbb{X}$ be the block column with entries $X_1, \dots, X_p$, and let $\mathbb{L} = [L_{\mu\nu}]$ be a $p \times p$ operator block matrix. In our normalization,

$$C^{(n)}_{\mathrm{NH}}(W) = n \inf_{\mathbb{L}, X} \mathrm{Tr}[(W \otimes \rho^{(n)})\mathbb{L}], \tag{S22}$$

We form an auxiliary Hermitian block matrix $\mathbb{M}(\mathbb{L}, X)$ from $\mathbb{L}$, $\mathbb{X}$, and the identity on $\mathcal{H}^{\otimes n}$. The constraints are

$$\begin{aligned} L_{\mu\nu} = L_{\nu\mu} = L^\dagger_{\mu\nu}, \qquad & \mathbb{M}(\mathbb{L}, X) := \begin{pmatrix} \mathbb{L} & \mathbb{X} \\ \mathbb{X}^\dagger & \mathbb{I}_{\mathcal{H}^{\otimes n}} \end{pmatrix} \geq 0, \\ \mathrm{Tr}(\rho^{(n)} X_\mu) = 0, \qquad & \mathrm{Tr}(D^{(n)}_\nu X_\mu) = \delta_{\mu\nu}. \end{aligned} \tag{S23}$$

The trace in Eq. (S22) includes the parameter and probe spaces. Since the lower-right block of $\mathbb{M}$ is the identity, $\mathbb{M}(\mathbb{L}, X) \geq 0$ is equivalent to $\mathbb{L} - \mathbb{X}\mathbb{X}^\dagger \geq 0$. This difference is called the Schur complement of the identity block, and the equivalence is the Schur-complement criterion [11]. A locally unbiased measurement gives the moments $X_\mu = \sum_k \widetilde{\theta}_\mu(k) M_k$ and $L_{\mu\nu} = \sum_k \widetilde{\theta}_\mu(k)\widetilde{\theta}_\nu(k) M_k$. For these moments, the block matrix in Eq. (S23) can be written as

$$\mathbb{M}(\mathbb{L}, X) = \sum_k \begin{pmatrix} \widetilde{\boldsymbol{\theta}}(k) \\ 1 \end{pmatrix} \begin{pmatrix} \widetilde{\boldsymbol{\theta}}(k)^T & 1 \end{pmatrix} \otimes M_k \geq 0.$$

This proves the upper inequality in Eq. (S4). Taking the probe expectation of $\mathbb{L} \geq \mathbb{X}\mathbb{X}^\dagger$ gives a real symmetric matrix above the $n$-copy complex covariance with entries $\mathrm{Tr}(\rho^{(n)} X_\mu X_\nu)$. The corresponding weighted trace is therefore bounded below by the Holevo bound of the $n$-copy model, proving the lower inequality after multiplication by $n$. These arguments recover the standard hierarchy in Eq. (S4) [2, 7].

We now turn to the three-parameter rank-two model of Eq. (S5) and evaluate its one-copy NH bound for $W = \mathbb{I}_3$. We first construct a feasible solution of the semidefinite program (SDP), which gives an upper bound on its minimum, and then prove a matching lower bound. Define $\gamma = \frac{2+\sqrt{2}}{8}$, $S = \sigma_z^{(r)} \otimes \mathbb{I}_t$, and $T = \mathbb{I}_r \otimes \sigma_x^{(t)}$. The optimal single-copy measurement constructed in Sec. II B has the operator moments

$$X^*_\mu = \frac{D_\mu}{2}, \qquad \mathbb{L}^* = \frac{\gamma}{2} \begin{pmatrix} \sqrt{2}\,\mathbb{I} & 0 & 0 \\ 0 & \mathbb{I} & T \\ 0 & T & \mathbb{I} \end{pmatrix}. \tag{S24}$$

They obey local unbiasedness since $\mathrm{Tr}(\rho_0 D_\mu) = 0$ and $\mathrm{Tr}(D_\nu D_\mu) = 2\delta_{\mu\nu}$. To check the positivity constraint, we can define the block column $\mathbb{F} = (\mathbb{I}, \mathrm{i}S/\sqrt{2}, \mathrm{i}ST/\sqrt{2})^T$, where the transpose acts only on the block indices. Direct multiplication gives $\mathbb{L}^* - \mathbb{X}^*\mathbb{X}^{*\dagger} = (\sqrt{2}/8)\mathbb{F}\mathbb{F}^\dagger \geq 0$. The objective is therefore

$$\mathrm{Tr}[(\mathbb{I}_3 \otimes \rho_0)\mathbb{L}^*] = \frac{\gamma}{2}(\sqrt{2} + 2) = \frac{3 + 2\sqrt{2}}{8}. \tag{S25}$$

Equation (S25) therefore establishes $C^{(1)}_{\mathrm{NH}} \leq (3 + 2\sqrt{2})/8$. It remains to exclude smaller objective values within the relaxed feasible set. We use a dual certificate, a positive-semidefinite matrix $\mathbb{Y}$ chosen so that its trace against any feasible $\mathbb{M}(\mathbb{L}, X)$ equals the objective minus a fixed constant. Let $E = |0\rangle\langle 1|_r \otimes \mathbb{I}_t$ and define

$$\mathbb{Y} = \begin{pmatrix} \mathbb{A} & \mathbb{B} \\ \mathbb{B}^\dagger & 2\gamma^2 Q \end{pmatrix}, \qquad \mathbb{A} = \frac{1}{2} \begin{pmatrix} P & \mathrm{i}P/\sqrt{2} & \mathrm{i}PT/\sqrt{2} \\ -\mathrm{i}P/\sqrt{2} & P & 0 \\ -\mathrm{i}PT/\sqrt{2} & 0 & P \end{pmatrix}, \qquad \mathbb{B} = -\gamma \begin{pmatrix} \mathrm{i}E \\ E/\sqrt{2} \\ ET/\sqrt{2} \end{pmatrix}. \tag{S26}$$

We first verify that $\mathbb{Y}$ is positive semidefinite. The matrix $\mathbb{A}$ vanishes on $\mathbb{C}^3 \otimes Q\mathcal{H}$. On each flag sector $T = a$, $a = \pm 1$, its restriction to $\mathbb{C}^3 \otimes P\mathcal{H}$ is

$$\frac{1}{2} \begin{pmatrix} 1 & \mathrm{i}/\sqrt{2} & \mathrm{i}a/\sqrt{2} \\ -\mathrm{i}/\sqrt{2} & 1 & 0 \\ -\mathrm{i}a/\sqrt{2} & 0 & 1 \end{pmatrix},$$

whose eigenvalues are 0, 1/2, and 1. Hence $\mathbb{A} \geq 0$. Direct multiplication gives $\mathbb{A}\mathbb{B} = \mathbb{B}$ and $\mathbb{B}^\dagger\mathbb{B} = 2\gamma^2 Q$, so

$$\mathbb{Y} = \begin{pmatrix} \mathbb{I}_{12} \\ \mathbb{B}^\dagger \end{pmatrix} \mathbb{A} \begin{pmatrix} \mathbb{I}_{12} & \mathbb{B} \end{pmatrix} \geq 0.$$

For any feasible pair, $\mathbb{M}(\mathbb{L}, X) \geq 0$, so $\mathrm{Tr}[\mathbb{Y}\mathbb{M}(\mathbb{L}, X)] \geq 0$. This inequality applies to the full SDP, including variables with off-diagonal blocks between flag sectors.

We next express this nonnegative trace in terms of the SDP objective and the unbiasedness constraints. The diagonal blocks of $\mathbb{A}$ equal $\rho_0$ and its off-diagonal blocks obey $A_{\mu\nu} + A_{\nu\mu} = 0$. The symmetry of $L_{\mu\nu}$ thus gives $\mathrm{Tr}(\mathbb{A}\mathbb{L}) = \mathrm{Tr}[(\mathbb{I}_3 \otimes \rho_0)\mathbb{L}]$, which is precisely the objective. The terms involving $X_\mu$ are fixed by local unbiasedness, since the blocks of $\mathbb{B}$ satisfy

$$B_1 + B_1^\dagger = -\sqrt{2}\gamma D_1, \qquad B_2 + B_2^\dagger = -\gamma D_2, \qquad B_3 + B_3^\dagger = -\gamma D_3.$$

Using local unbiasedness and $\mathrm{Tr}\, Q = 2$ therefore yields

$$\begin{aligned} 0 \leq \mathrm{Tr}[\mathbb{Y}\mathbb{M}(\mathbb{L}, X)] &= \mathrm{Tr}[(\mathbb{I}_3 \otimes \rho_0)\mathbb{L}] - (\sqrt{2} + 2)\gamma + 4\gamma^2, \\ C_{\mathrm{NH}}^{(1)} &\geq (\sqrt{2} + 2)\gamma - 4\gamma^2 = \frac{3 + 2\sqrt{2}}{8}. \end{aligned} \tag{S27}$$

The lower bound in Eq. (S27) matches the feasible value in Eq. (S25). The moments in Eq. (S24) therefore minimize the NH SDP, whose value equals the single-copy optimum already established in Eq. (S21):

$$C_{\mathrm{NH}}^{(1)} = C^{(1)} = \frac{3 + 2\sqrt{2}}{8}. \tag{S28}$$

The two-copy NH value follows from the bound hierarchy and an attaining measurement. As we will show in Sec. II D, the two-copy measurement attains the normalized Holevo value 5/8. The lower and upper bounds on $C_{\mathrm{NH}}^{(2)}$ in Eq. (S4) then coincide, forcing $C_{\mathrm{NH}}^{(2)} = C^{(2)} = C^{\mathrm{H}} = 5/8$. Combining this with the one-copy value gives

$$C_{\mathrm{NH}}^{(1)} - C^{\mathrm{H}} = \frac{\sqrt{2} - 1}{4} > 0, \qquad C_{\mathrm{NH}}^{(2)} - C^{\mathrm{H}} = 0. \tag{S29}$$

This establishes failure of weak gap persistence between the Nagaoka–Hayashi and Holevo bounds for three parameters [3]. The most informative and Nagaoka–Hayashi bounds coincide at both copy numbers, so this example does not resolve the separate persistence question for the gap between them.

### D. Two-copy projective measurement

Having established the strict single-copy gap, we now construct a two-copy projective measurement and a locally unbiased estimator with normalized cost equal to the Holevo bound $C^{\mathrm{H}} = 5/8$. We specify the measurement through commuting estimator observables, then verify local unbiasedness and compute its cost.

We use the ordering $(r_1 r_2) \otimes (t_1 t_2)$ for the two-copy Hilbert space. Measuring $t_1$ and $t_2$ in their joint $\sigma_x$ eigenbasis $|a_x, b_x\rangle$ gives classical flags $a, b \in \{+1, -1\}$. The subsequent measurement on $r_1 r_2$ depends on these flags. To express its observables using Pauli matrices, we introduce the ordered basis

$$(|s\rangle, |u_+\rangle, |u_-\rangle, |11\rangle), \qquad |s\rangle = |00\rangle, \qquad |u_\pm\rangle = \frac{|10\rangle \pm |01\rangle}{\sqrt{2}}. \tag{S30}$$

Let $U_r$ be the unitary that maps the computational basis of two effective qubits to this ordered basis,

$$U_r|00\rangle = |s\rangle, \quad U_r|01\rangle = |u_+\rangle, \quad U_r|10\rangle = |u_-\rangle, \quad U_r|11\rangle = |11\rangle.$$

Within each flag sector, we choose Hermitian estimator observables whose joint eigenvalues will give the three parameter estimates. For equal flags, $ab = +1$, define

$$X_1^{(2)}(a, b) = 0, \quad X_2^{(2)}(a, b) = U_r \left(\frac{\mathbb{I} \otimes \sigma_x}{2}\right) U_r^\dagger, \quad X_3^{(2)}(a, b) = \frac{a}{2} X_2^{(2)}(a, b). \tag{S31}$$

For opposite flags, $ab = -1$, define

$$X_1^{(2)}(a,b) = U_r\left(-\frac{\mathbb{I}\otimes\sigma_y}{2}\right)U_r^\dagger, \quad X_2^{(2)}(a,b) = 0, \quad X_3^{(2)}(a,b) = aU_r\left(\frac{\sigma_x\otimes\mathbb{I}}{4}\right)U_r^\dagger. \tag{S32}$$

Combining these conditional observables over the four orthogonal flag sectors gives the full observables

$$X_\mu^{(2)} = \sum_{a,b} X_\mu^{(2)}(a,b)\otimes|a_x,b_x\rangle\langle a_x,b_x|. \tag{S33}$$

For $ab = +1$, the two nonzero observables are proportional. For $ab = -1$, they act on different effective tensor factors before the common basis change $U_r$. Thus the observables commute within each sector. Their block-diagonal form in Eq. (S33) then gives

$$[X_\mu^{(2)}, X_\nu^{(2)}] = 0 \tag{S34}$$

for all $\mu,\nu$. A common eigenbasis therefore defines a projective measurement, with the joint eigenvalues used as the parameter estimates.

To check local unbiasedness, we now evaluate the means and their first derivatives for this measurement. At the estimation point, the two-copy state is

$$\rho_0^{\otimes 2} = |s\rangle\langle s|\otimes\frac{\mathbb{I}_{t_1t_2}}{4}.$$

The parameter derivatives are determined by the two-copy tangent operators $D_\mu^{(2)}$. Within a flag sector $(a,b)$, their components connecting the support to its orthogonal complement take $|s\rangle$ to the following vectors for $\mu = 1,2,3$, respectively:

$$-\frac{\mathrm{i}}{2}|u_+\rangle, \qquad \frac{1}{2}|u_+\rangle, \qquad \begin{cases}\frac{a}{2}|u_+\rangle, & ab=+1,\\ \frac{a}{2}|u_-\rangle, & ab=-1.\end{cases} \tag{S35}$$

These are blocks of the full tangent operators, including the sector probability 1/4. Substituting the observables of Eqs. (S31) and (S32) and summing over all four sectors gives

$$\mathrm{Tr}(\rho_0^{\otimes 2}X_\mu^{(2)}) = 0, \qquad \mathrm{Tr}(D_\nu^{(2)}X_\mu^{(2)}) = \delta_{\mu\nu}. \tag{S36}$$

The first identity in Eq. (S36) gives zero mean, and the second gives the identity response matrix. These are precisely the local unbiasedness conditions for the estimator. For the joint measurement, its mean-square error matrix is obtained from the second moments of the commuting observables,

$$V_{\mu\nu}^{(2)} = \mathrm{Tr}(\rho_0^{\otimes 2}X_\mu^{(2)}X_\nu^{(2)}) = \mathrm{diag}\left(\frac{1}{8},\frac{1}{8},\frac{1}{16}\right)_{\mu\nu}. \tag{S37}$$

Thus the measurement achieves the normalized cost $2\,\mathrm{Tr}\,V^{(2)} = 5/8$. Additivity of the Holevo bound gives $\mathrm{Tr}\,V^{(2)} \geq 5/16$ for any locally unbiased two-copy measurement, so the constructed measurement is globally optimal. This establishes $C^{(2)} = C^{\mathrm{H}} = 5/8$ and closes the strict single-copy gap.

To specify the measurement explicitly, we now list a common eigenbasis and the estimate assigned to each outcome. Let $|z\rangle$ be a computational-basis state, let $|\lambda_x\rangle$ be a $\sigma_x$ eigenstate with eigenvalue $\lambda$, and let $|\eta_y\rangle$ be a $\sigma_y$ eigenstate with eigenvalue $\eta$. When $ab = +1$, the joint measurement can be refined to the four rank-one projectors

$$\Pi_{abz\lambda}^{(+)} = U_r\left(|z\rangle\langle z|\otimes|\lambda_x\rangle\langle\lambda_x|\right)U_r^\dagger\otimes|a_x,b_x\rangle\langle a_x,b_x|, \tag{S38}$$

where $z\in\{0,1\}$ and $\lambda\in\{+1,-1\}$. Assign the estimator value

$$\widetilde{\boldsymbol{\theta}}_{abz\lambda}^{(+)} = \left(0,\frac{\lambda}{2},\frac{a\lambda}{4}\right)^T.$$

When $ab = -1$, use

$$\Pi_{ab\lambda\eta}^{(-)} = U_r\left(|\lambda_x\rangle\langle\lambda_x|\otimes|\eta_y\rangle\langle\eta_y|\right)U_r^\dagger\otimes|a_x,b_x\rangle\langle a_x,b_x|, \tag{S39}$$

where $\lambda, \eta \in \{+1, -1\}$. Assign

$$\widetilde{\boldsymbol{\theta}}^{(-)}_{ab\lambda\eta} = \left(-\frac{\eta}{2}, 0, \frac{a\lambda}{4}\right)^T.$$

The sixteen projectors in Eqs. (S38) and (S39) are mutually orthogonal and sum to the identity. The listed estimates are the joint eigenvalues of the observables in Eq. (S33). These projectors and estimates therefore give an explicit four-qubit realization of the locally unbiased measurement with the covariance in Eq. (S37).

### E. Single-copy obstruction

To understand why a single copy cannot attain the Holevo bound, we examine the commutativity constraints on estimator observables in a Naimark dilation. Assume that a single-copy measurement and locally unbiased estimator attain 5/8. In a Naimark dilation [1], the estimator is represented by commuting Hermitian observables $\widetilde{X}_\mu$. These observables must also satisfy the Holevo equality conditions. We will show that the two requirements are incompatible. Let $P$ now denote the state-support projector embedded in the dilation space, and define the total support-to-complement maps

$$\widetilde{\mathcal{B}}_\mu = (\mathbb{I} - P)\widetilde{X}_\mu P.$$

Because the observables commute, their Holevo functional equals the measurement cost. A cost of 5/8 therefore forces equality in the lower-bound chain of Eq. (S12). Applying the equality family of Eq. (S14) in the dilation space gives

$$\widetilde{\mathcal{B}}_1 = -\frac{\mathrm{i}}{2}\Gamma + R, \qquad \widetilde{\mathcal{B}}_2 = \frac{1}{2}\Gamma - \mathrm{i}R, \qquad \widetilde{\mathcal{B}}_3 = \frac{1}{2}\Gamma_x, \tag{S40}$$

with the orthogonality conditions in Eq. (S15). The support blocks $P\widetilde{X}_\mu P$ vanish.

We now impose commutativity on this equality family. Since the support blocks vanish, the support block of each commutator depends only on the maps $\widetilde{\mathcal{B}}_\mu$. Compressing $[\widetilde{X}_\mu, \widetilde{X}_\nu] = 0$ to the state support gives

$$\widetilde{\mathcal{B}}_\mu^\dagger \widetilde{\mathcal{B}}_\nu - \widetilde{\mathcal{B}}_\nu^\dagger \widetilde{\mathcal{B}}_\mu = 0.$$

For $(\mu, \nu) = (1, 2)$, substituting Eq. (S40) and using $\Gamma^\dagger\Gamma = \mathbb{I}_t/2$ yields

$$R^\dagger R = \frac{1}{8}\mathbb{I}_t. \tag{S41}$$

Equation (S41) fixes $R^\dagger R$ for a putative saturating measurement. The remaining pairs $(1, 3)$ and $(2, 3)$ constrain the overlap of $R$ with $\Gamma_x$, giving, respectively,

$$R^\dagger\Gamma_x - \Gamma_x^\dagger R = -\frac{\mathrm{i}}{2}\sigma_x^{(t)}, \quad R^\dagger\Gamma_x + \Gamma_x^\dagger R = 0.$$

Combining these two equations gives

$$R^\dagger\Gamma_x = -\frac{\mathrm{i}}{4}\sigma_x^{(t)}. \tag{S42}$$

To check whether additional Naimark directions can satisfy these constraints, decompose $R = R_Q + R_\perp$, where the range of $R_Q$ lies in the physical kernel $Q\mathcal{H}$ and the range of $R_\perp$ lies in the additional Naimark space. Since $\Gamma_x$ has range only in the physical kernel, its overlap with $R$ depends only on $R_Q$, and Eq. (S42) becomes

$$R_Q^\dagger \frac{\sigma_x^{(t)}}{\sqrt{2}} = -\frac{\mathrm{i}}{4}\sigma_x^{(t)}.$$

Multiplication by $\sigma_x^{(t)}$ gives $R_Q = \frac{\mathrm{i}}{2}\Gamma, \quad R_Q^\dagger R_Q = \frac{1}{8}\mathbb{I}_t$. The two ranges are orthogonal, so $R^\dagger R = R_Q^\dagger R_Q + R_\perp^\dagger R_\perp$. The physical-kernel component already accounts for the full $R^\dagger R$ required by Eq. (S41). Thus $R_\perp^\dagger R_\perp = 0$ and $R_\perp = 0$. We conclude that $R = \mathrm{i}\Gamma/2$, which gives

$$\langle \Gamma, R\rangle_{\mathrm{HS}} = \frac{\mathrm{i}}{2} \neq 0. \tag{S43}$$

This violates the required orthogonality $\langle \Gamma, R\rangle_{\mathrm{HS}} = 0$ in Eq. (S15). Therefore no locally unbiased single-copy POVM attains the Holevo bound. The contradiction identifies the commutativity obstruction underlying the strict gap quantified independently by Eq. (S21).

## III. FULL-RANK PERSISTENCE THEOREM

We prove the full-rank persistence theorem by showing that exact Holevo saturation with any finite number of copies implies saturation with one copy for full-rank models. We assume $\rho > 0$. An invertible real linear reparameterization reduces any real symmetric positive-definite weight matrix $W = W^T > 0$ to the identity and transforms the local unbiasedness constraints accordingly, so we take $W = \mathbb{I}_p$ throughout the proof with $p$ being the number of estimated parameters. We first establish that the measurement optimum is attained. We then show that saturation forces the compressed estimator operators to commute, and use uniqueness of the Holevo optimizer to transfer this commutativity from $n$ copies to one.

Attainment is needed to distinguish a strictly positive gap from an unattained infimum equal to the Holevo bound. We begin with one copy. By Lemma 3 of Ref. [2], which rests on Carathéodory's theorem, restricting the minimization to measurements with at most $m = p(p+1)(\dim\mathcal{H})^2/2+1$ outcomes does not change its value. The set of $m$-outcome measurements is compact, so it suffices to show that the minimal locally unbiased cost is lower semicontinuous on this set.

To show that local unbiasedness is preserved in the limit, consider a convergent sequence of $m$-outcome measurements with locally unbiased estimators whose costs converge to a finite value $c$. Write $p_k = \mathrm{Tr}(\rho M_k)$ for the outcome probabilities of a member of the sequence and $\|\cdot\|_{\mathrm{op}}$ for the operator norm. Since $\rho > 0$, each element obeys $\mathrm{Tr}(M_k) \le p_k/\lambda_{\min}(\rho)$, where $\lambda_{\min}(\rho)$ is the smallest eigenvalue of $\rho$, so each term of an unbiasedness sum satisfies

$$\big|\,\mathrm{Tr}(D_\nu M_k)\,\widetilde{\theta}_\mu(k)\big| \le \frac{\|D_\nu\|_{\mathrm{op}}}{\lambda_{\min}(\rho)}\sqrt{p_k}\,\sqrt{p_k\,|\widetilde{\boldsymbol{\theta}}(k)|^2},$$

Replacing $D_\nu$ by $\rho$ in this estimate controls the mean condition. The last factor is bounded along the sequence because $p_k|\widetilde{\boldsymbol{\theta}}(k)|^2 \le \mathrm{Tr}\,V$. Outcomes whose probabilities vanish in the limit therefore drop out of the unbiasedness sums, while for the remaining outcomes the estimator values are bounded and converge along a subsequence. The limit measurement with the limit estimator is locally unbiased with cost at most $c$. This proves lower semicontinuity and hence existence of a minimizer. For $n$ copies, the argument applies on $\mathcal{H}^{\otimes n}$ with $\rho^{\otimes n} > 0$ and $\dim\mathcal{H}$ replaced by $(\dim\mathcal{H})^n$ in the outcome bound. Thus $C^{(n)} = C^{\mathrm{H}}$ guarantees an actual saturating measurement.

### A. Minimal Naimark noise

We first determine what Holevo saturation requires of the extra covariance introduced by a Naimark dilation. In this and the next subsection, we consider a general full-rank local model $(\rho, D_1, \ldots, D_p)$. The conclusions apply to the $n$-copy model by replacing $\rho$ and $D_\mu$ with $\rho^{\otimes n}$ and $D_\mu^{(n)}$, whose Holevo bound is $C^{\mathrm{H}}/n$.

Suppose a locally unbiased measurement attains the Holevo bound of the model under consideration. In its Naimark representation with isometry $\mathcal{V}$, write $X_\mu = \mathcal{V}^\dagger \widetilde{X}_\mu \mathcal{V}$ for the compressed estimator operators and $E_\mu = (\mathbb{I} - \mathcal{V}\mathcal{V}^\dagger)\widetilde{X}_\mu \mathcal{V}$ for their components mapping into the added dilation space. The measurement covariance decomposes as

$$V = Z(X) + N, \qquad N_{\mu\nu} = \mathrm{Tr}(\rho E_\mu^\dagger E_\nu).$$

Thus $N$ is a positive-semidefinite Gram matrix of the maps $E_\mu\sqrt{\rho}$, and its real part is the added covariance. Since $V$ is real symmetric, the imaginary part of $N$ must cancel that of $Z$. Writing $\Omega = \mathrm{Im}\,Z$ and $S = \mathrm{Re}\,N$, we have $N = S - \mathrm{i}\Omega$ with $S = S^T \in \mathbb{R}^{p\times p}$. Both $S - \mathrm{i}\Omega$ and its complex conjugate $S + \mathrm{i}\Omega$ are positive semidefinite.

The following matrix fact bounds the added cost $\mathrm{Tr}\,S$ and characterizes equality. If $\Omega$ is real antisymmetric and $S \pm \mathrm{i}\Omega \ge 0$, then

$$\mathrm{Tr}\,S \ge \|\Omega\|_1\,. \tag{S44}$$

Here $|\mathrm{i}\Omega| = \sqrt{(\mathrm{i}\Omega)^2}$. At equality the unique solution is

$$S = |\mathrm{i}\Omega|, \qquad N = |\mathrm{i}\Omega| - \mathrm{i}\Omega. \tag{S45}$$

To prove this statement, let $\Pi_+$, $\Pi_-$, and $\Pi_0$ be the positive, negative, and zero spectral projectors of the Hermitian matrix $\mathrm{i}\Omega$. Positivity of $S - \mathrm{i}\Omega$ on $\Pi_+$ and of $S + \mathrm{i}\Omega$ on $\Pi_-$ gives

$$\mathrm{Tr}(\Pi_+ S) \ge \mathrm{Tr}(\Pi_+ \mathrm{i}\Omega), \qquad \mathrm{Tr}(\Pi_- S) \ge -\,\mathrm{Tr}(\Pi_- \mathrm{i}\Omega).$$

The compression $\Pi_0 S \Pi_0$ is positive semidefinite. Summing the three sectors proves Eq. (S44). At equality, the positive-semidefinite matrices $S - \mathrm{i}\Omega$, $S + \mathrm{i}\Omega$, and $S$ have vanishing compressions on $\Pi_+$, $\Pi_-$, and $\Pi_0$, respectively. A positive-semidefinite matrix with a zero diagonal block also has zero off-diagonal blocks connected to it. These conditions determine $S$ on all three sectors and give Eq. (S45).

Returning to the measurement cost, $\operatorname{Tr} V = \operatorname{Tr}\operatorname{Re} Z + \operatorname{Tr} S$, so Eq. (S44) gives

$$\operatorname{Tr} V \geq \operatorname{Tr}\operatorname{Re} Z + \|\operatorname{Im} Z\|_1 \geq C^{\mathrm{H}}(\mathbb{I}_p). \tag{S46}$$

Exact measurement saturation forces equality in both steps. The second equality makes $X$ a Holevo optimizer, while the first fixes its leakage Gram matrix to Eq. (S45). The Gram condition in Eq. (S45) must therefore be compatible with the commutativity of the Naimark observables.

### B. Full rank removes nonzero imaginary covariance

We now show that the minimal-noise condition is incompatible with $\Omega \neq 0$ when $\rho > 0$. A real orthogonal change of parameter coordinates preserves $W = \mathbb{I}_p$ and puts $\Omega$ into $2 \times 2$ skew blocks and a zero block. If $\Omega \neq 0$, we can choose a nonzero block of the form

$$\Omega\big|_{\{1,2\}} = \kappa \begin{pmatrix} 0 & 1 \\ -1 & 0 \end{pmatrix}, \qquad \kappa > 0. \tag{S47}$$

Equation (S45) fixes the corresponding Gram block to $N|_{\{1,2\}} = \kappa \left(\begin{smallmatrix} 1 & -\mathrm{i} \\ \mathrm{i} & 1 \end{smallmatrix}\right)$. Its rank-one form requires $E_2\sqrt{\rho} = -\mathrm{i}E_1\sqrt{\rho}$. Here full rank is essential, since invertibility of $\sqrt{\rho}$ turns this state-weighted relation into the operator identity $E_2 = -\mathrm{i}E_1$. Compression of $[\widetilde{X}_1, \widetilde{X}_2] = 0$ to the original Hilbert space yields

$$[X_1, X_2] + E_1^\dagger E_2 - E_2^\dagger E_1 = 0.$$

Substitution gives

$$\mathrm{i}[X_1, X_2] = -2E_1^\dagger E_1 \leq 0. \tag{S48}$$

The ordinary trace of a finite-dimensional commutator vanishes. Taking the trace of Eq. (S48) therefore gives $E_1 = 0$, whereas the Gram block requires $\operatorname{Tr}(\rho E_1^\dagger E_1) = \kappa > 0$. This contradiction excludes all nonzero skew blocks, so $\Omega = 0$. Equation (S45) then gives $N = 0$. Full rank forces all $E_\mu = 0$, and compression of the commuting Naimark observables leaves commuting operators $X_\mu$. Thus exact saturation for a full-rank model, including any fixed finite-copy model, requires

$$\operatorname{Im} Z = 0, \qquad N = 0, \qquad [X_\mu, X_\nu] = 0. \tag{S49}$$

### C. Uniqueness and reduction to one copy

We have shown that a saturating measurement gives a commuting Holevo optimizer. To relate this optimizer on $n$ copies to a one-copy optimizer, we first prove uniqueness. For $W = \mathbb{I}_p$, the variational expression for the trace norm writes the Holevo functional as

$$h(X) = \max_K h_K(X), \qquad h_K(X) = \operatorname{Tr}[(\mathbb{I}_p + \mathrm{i}K)Z(X)],$$

where $K$ ranges over real antisymmetric matrices with $\|K\|_{\mathrm{op}} \leq 1$. Then $\mathbb{I}_p + \mathrm{i}K \geq 0$. For fixed $K$, $h_K$ is a nonnegative quadratic form in the Hermitian operators $X_\mu$. We will show that full rank makes each of these forms positive definite, which yields strict convexity of $h$.

For a set of Hermitian operators $\Delta$ with $h_K(\Delta) = 0$, a Gram factorization and $\rho > 0$ give $\sum_\mu (\mathbb{I}_p + \mathrm{i}K)_{\mu\nu}\Delta_\mu = 0$ for each $\nu$. Taking the Hermitian adjoint replaces $\mathrm{i}K$ by $-\mathrm{i}K$. Adding the two equations gives $2\Delta_\nu = 0$. Thus $h_K$ vanishes only when all operators in $\Delta$ are zero and is positive definite on sets of Hermitian operators, even if $\mathbb{I}_p + \mathrm{i}K$ itself is singular.

To pass from these quadratic forms to $h$, we can take distinct feasible sets of estimator operators $X$ and $X'$ and choose $K_*$ that attains the maximum at their midpoint. The quadratic identity gives

$$h\left(\frac{X+X'}{2}\right) = \frac{1}{2}h_{K_*}(X) + \frac{1}{2}h_{K_*}(X') - \frac{1}{4}h_{K_*}(X-X') < \frac{1}{2}h(X) + \frac{1}{2}h(X'). \tag{S50}$$

The strict inequality follows from $h_{K_*}(X-X') > 0$. Hence the Holevo functional is strictly convex on the affine space of locally unbiased Hermitian estimator operators. A minimizer exists because $h(X) \geq \lambda_{\min}(\rho)\sum_\mu \|X_\mu\|_{\mathrm{HS}}^2$ bounds the estimator operators at finite cost and the feasible set is closed. Denote the unique one-copy optimizer by $X^{\mathrm{H}}$.

We next identify the $n$-copy optimizer by averaging $X^{\mathrm{H}}$ over the copies. For a single-copy operator $O$, let $O^{[m]}$ denote its action on copy $m$ and the identity on all other copies. We define the additive estimator operators

$$X_\mu^{\mathrm{H},(n)} = \frac{1}{n}\sum_{m=1}^{n}(X_\mu^{\mathrm{H}})^{[m]} \tag{S51}$$

The product-state form and the one-copy unbiasedness conditions make these estimator operators locally unbiased. Their cross-copy second moments vanish because $\mathrm{Tr}(\rho X_\mu^{\mathrm{H}}) = 0$, leaving the complex covariance $Z(X^{\mathrm{H}})/n$. Their Holevo functional therefore equals $C^{\mathrm{H}}(\mathbb{I}_p)/n$, the $n$-copy Holevo bound. Since $\rho^{\otimes n}$ remains full rank, strict convexity makes Eq. (S51) the unique $n$-copy optimizer.

We can now transfer commutativity to one copy. If an $n$-copy measurement attains the Holevo bound, its compressed estimator operators must coincide with those in Eq. (S51) by uniqueness and must commute by Eq. (S49). Operators acting on different copies commute, so

$$0 = [X_\mu^{\mathrm{H},(n)}, X_\nu^{\mathrm{H},(n)}] = \frac{1}{n^2}\sum_{r=1}^{n}[X_\mu^{\mathrm{H}}, X_\nu^{\mathrm{H}}]^{[r]}.$$

For $n > 1$, the ordinary partial trace over copies $2,\ldots,n$ removes each term with $r > 1$ because the trace of a commutator vanishes. The surviving term is $(\dim\mathcal{H})^{n-1}/n^2$ times the one-copy commutator. Thus, also including the immediate case $n = 1$,

$$[X_\mu^{\mathrm{H}}, X_\nu^{\mathrm{H}}] = 0. \tag{S52}$$

The joint spectral measurement of these commuting one-copy operators is locally unbiased and has cost equal to their Holevo functional. It therefore attains the one-copy Holevo bound. We have proved that finite-copy saturation implies one-copy saturation. Since the measurement optimum is attained at each finite copy number, as established at the start of this section, a strict one-copy gap rules out equality at any finite $n$. Restoring the original weight matrix gives

$$\rho > 0, \qquad C^{(1)}(W) > C^{\mathrm{H}}(W) \implies C^{(n)}(W) > C^{\mathrm{H}}(W) \tag{S53}$$

for any $W = W^T > 0$, any parameter number, and each integer $n \geq 1$.

The theorem also covers rank-deficient local models whose support is stationary to first order, because their local estimation problem reduces to a full-rank problem on the support. To see this, let $P$ be the support projector and assume $(\mathbb{I}-P)D_\mu P = 0$ for all $\mu$. Hermiticity also gives $PD_\mu(\mathbb{I}-P) = 0$. Smooth positivity for both signs of a parameter displacement gives $(\mathbb{I}-P)D_\mu(\mathbb{I}-P) = 0$. Indeed, along the coordinate line $\theta_\mu = t$ with all other parameters fixed at zero, the expectation of the state in each kernel vector is a nonnegative differentiable function of $t$ with a minimum at $t = 0$, so its derivative vanishes. The polarization identity then removes the full kernel block. Hence $D_\mu = PD_\mu P$, so both the state and all first-order parameter dependence lie in $P\mathcal{H}$. Compressing a measurement to this subspace preserves its local outcome probabilities and derivatives. Since $\rho$ is full rank on $P\mathcal{H}$, Eq. (S53) applies there. Finite-copy closure of a strict single-copy gap therefore requires both rank deficiency and first-order support motion. The model in Eq. (S5) satisfies both conditions.

## IV. EXACT BOUNDS FOR THE DEPOLARIZED MODEL

We now consider noisy states with controlled depolarization:

$$\rho_{\boldsymbol{\theta},\epsilon} = (1-\epsilon)\rho_{\boldsymbol{\theta}} + \epsilon\frac{\mathbb{I}}{4}, \qquad 0 \leq \epsilon < 1, \tag{S54}$$

where $\rho_{\boldsymbol{\theta}}$ is the model of Eq. (S5). The family is smooth and locally identifiable, its tangent operators are $(1-\epsilon)D_\mu$, and

$$\rho_{\mathbf{0},\epsilon} = \frac{2-\epsilon}{4}P + \frac{\epsilon}{4}Q.$$

It is full rank for each $\epsilon > 0$. We write $C^{(n)}(\epsilon)$ and $C^{\mathrm{H}}(\epsilon)$ for the quantities of Eqs. (S1) and (S3) evaluated on this family with $W = \mathbb{I}_3$.

The single-copy optimum follows from the classical-quantum decomposition. Depolarizing preserves the block structure of Eq. (S17), so conditioning on $a$ gives the qubit state $(1-\epsilon)\rho^{(r)}_{\boldsymbol{\theta},a} + \epsilon\,\mathbb{I}/2$. Its Bloch vector keeps the constant length $1-\epsilon$ and rotates exactly as in the noiseless conditional model. For a qubit family whose Bloch vector $\boldsymbol{r}$ has constant length, the quantum Fisher information matrix is $J_{\mu\nu} = \partial_\mu \boldsymbol{r}\cdot\partial_\nu \boldsymbol{r}$, so the conditional matrix becomes $J_a = 8(1-\epsilon)^2\mathbb{I}_2$. The measurement reduction, the Gill–Massar inequality, and the symmetrization argument of Sec. II B apply without change, with the constraint $x+y \le 8(1-\epsilon)^2$. We use the same randomized $\sigma_y^{(r)}$ and $\sigma_x^{(r)}$ measurements as in the noiseless model. Depolarization reduces their Fisher information by a factor of $(1-\epsilon)^2$, and the local score estimator attains the resulting classical Cramér–Rao bound. Therefore we have

$$C^{(1)}(\epsilon) = \frac{3+2\sqrt{2}}{8(1-\epsilon)^2}. \tag{S55}$$

The Holevo bound of the depolarized model is also exact. For locally unbiased estimator operators $X_\mu$, define $\mathcal{A}_\mu$ and $\mathcal{B}_\mu$ as before and let $\overline{\mathcal{A}}_\mu = QX_\mu Q$. Write

$$Z_{\epsilon,\mu\nu}(X) = \mathrm{Tr}(\rho_{\mathbf{0},\epsilon}X_\mu X_\nu), \qquad h_\epsilon(X) = \mathrm{Tr}\,\mathrm{Re}\,Z_\epsilon(X) + \|\mathrm{Im}\,Z_\epsilon(X)\|_1\,,$$

and introduce the quadratic expression

$$\begin{aligned} q_\epsilon(X) = {} & \mathrm{Tr}[\rho_{\mathbf{0},\epsilon}(X_1+\mathrm{i}X_2)(X_1-\mathrm{i}X_2)] \\ & + \mathrm{Tr}(\rho_{\mathbf{0},\epsilon}X_3^2). \end{aligned}$$

The trace-norm inequality gives $h_\epsilon(X) \ge q_\epsilon(X)$. Using the support-kernel decomposition above, block multiplication gives

$$\begin{aligned} q_\epsilon(X) = {} & \frac{2-\epsilon}{4}\left(\|\mathcal{A}_1-\mathrm{i}\mathcal{A}_2\|_{\mathrm{HS}}^2 + \|\mathcal{B}_1-\mathrm{i}\mathcal{B}_2\|_{\mathrm{HS}}^2 + \|\mathcal{A}_3\|_{\mathrm{HS}}^2 + \|\mathcal{B}_3\|_{\mathrm{HS}}^2\right) \\ & + \frac{\epsilon}{4}\left(\|\mathcal{B}_1+\mathrm{i}\mathcal{B}_2\|_{\mathrm{HS}}^2 + \left\|\overline{\mathcal{A}}_1-\mathrm{i}\overline{\mathcal{A}}_2\right\|_{\mathrm{HS}}^2 + \|\mathcal{B}_3\|_{\mathrm{HS}}^2 + \left\|\overline{\mathcal{A}}_3\right\|_{\mathrm{HS}}^2\right). \end{aligned} \tag{S56}$$

The local unbiasedness conditions are those of Eq. (S10) multiplied by $1-\epsilon$, so

$$\langle\Gamma,\mathcal{B}_1-\mathrm{i}\mathcal{B}_2\rangle_{\mathrm{HS}} = -\frac{\mathrm{i}}{1-\epsilon}, \qquad \mathrm{Re}\langle\Gamma_x,\mathcal{B}_3\rangle_{\mathrm{HS}} = \frac{1}{2(1-\epsilon)}.$$

Cauchy-Schwarz and the omitted nonnegative terms of Eq. (S56) give

$$h_\epsilon(X) \ge \frac{2-\epsilon}{4(1-\epsilon)^2} + \frac{1}{8(1-\epsilon)^2} = \frac{5-2\epsilon}{8(1-\epsilon)^2}.$$

The bound is attained by choosing $\mathcal{A}_\mu = \overline{\mathcal{A}}_\mu = 0$ and

$$\mathcal{B}_1 = -\frac{\mathrm{i}\Gamma}{2(1-\epsilon)}, \qquad \mathcal{B}_2 = \frac{\Gamma}{2(1-\epsilon)}, \qquad \mathcal{B}_3 = \frac{\Gamma_x}{2(1-\epsilon)}.$$

Direct evaluation gives $\mathrm{Tr}\,\mathrm{Re}\,Z_\epsilon = 3/[8(1-\epsilon)^2]$ and $\|\mathrm{Im}\,Z_\epsilon\|_1 = 1/[4(1-\epsilon)]$. Hence

$$C^{\mathrm{H}}(\epsilon) = \frac{5-2\epsilon}{8(1-\epsilon)^2}. \tag{S57}$$

Subtracting Eq. (S57) from Eq. (S55) gives the exact single-copy gap

$$C^{(1)}(\epsilon) - C^{\mathrm{H}}(\epsilon) = \frac{\sqrt{2}-1+\epsilon}{4(1-\epsilon)^2} > 0, \qquad 0 \le \epsilon < 1. \tag{S58}$$

We next evaluate the fixed two-copy projection-valued measurement (PVM) of Eqs. (S38) and (S39) on the depolarized states. The measurement is unchanged, while its locally optimal estimator is the score estimator of its classical outcome distribution. Let $p_j$ denote the probabilities of its 16 outcomes and $I_\epsilon^{\mathrm{PVM}}$ the classical Fisher information matrix at $\boldsymbol{\theta} = 0$, with

$$[I_\epsilon^{\mathrm{PVM}}]_{\mu\nu} = \sum_j \frac{\partial_\mu p_j \, \partial_\nu p_j}{p_j}.$$

Direct evaluation gives

$$I_\epsilon^{\mathrm{PVM}} = 2(1-\epsilon)^2 \operatorname{diag}\big((2-\epsilon)^2,\ 3+(1-\epsilon)^2,\ (2-\epsilon)(4-\epsilon)\big)\,. \tag{S59}$$

The corresponding normalized cost is

$$\begin{aligned} C_{\mathrm{PVM}}^{(2)}(\epsilon) &= 2\,\mathrm{Tr}[(I_\epsilon^{\mathrm{PVM}})^{-1}] \\ &= \frac{1}{(1-\epsilon)^2}\left[\frac{1}{(2-\epsilon)^2} + \frac{1}{3+(1-\epsilon)^2} + \frac{1}{(2-\epsilon)(4-\epsilon)}\right]. \end{aligned} \tag{S60}$$

This is an achievable upper bound on the exact two-copy optimum,

$$C^{(2)}(\epsilon) \le C_{\mathrm{PVM}}^{(2)}(\epsilon),$$

with equality to the Holevo value at $\epsilon = 0$. The fixed PVM beats the exact single-copy optimum for $0 \le \epsilon < \epsilon_*$, where the unique crossover is

$$C_{\mathrm{PVM}}^{(2)}(\epsilon_*) = C^{(1)}(\epsilon_*), \qquad \epsilon_* = 0.198262\ldots. \tag{S61}$$

Each term in the brackets of Eq. (S60) increases strictly with $\epsilon$. Their sum rises from $5/8 < (3+2\sqrt{2})/8$ at $\epsilon = 0$ to $5/3 > (3+2\sqrt{2})/8$ as $\epsilon \to 1^-$, so the crossover is unique.

For each $\epsilon > 0$, the depolarized family is full rank and Eq. (S58) gives a positive single-copy gap. The persistence theorem therefore applies, and the finite-copy gap persists at each integer $n \ge 1$. Exact finite-copy saturation along this family occurs only at the rank-deficient point $\epsilon = 0$, where two copies suffice. Equations (S55), (S57), and (S60) provide exact reference values for an experiment with controlled injected noise.

## V. EXPERIMENTAL IMPLEMENTATION AND DATA ANALYSIS

### A. Circuit construction and device

The classical-quantum decomposition of Eq. (S17) reduces each four-dimensional copy to one physical qubit and one classical bit. In the ideal model, the classical bit $a = \pm 1$ is the outcome of measuring $\sigma_x^{(t)}$, and its distribution is uniform and independent of $\boldsymbol{\theta}$. The joint statistics of the flags and the subsequent $r$ measurements are therefore reproduced exactly by drawing $(a, b)$ uniformly at random, recording them, preparing the conditional states $\rho_{\boldsymbol{\theta},a}^{(r)}$ and $\rho_{\boldsymbol{\theta},b}^{(r)}$ of Eq. (S18) on two qubits, and selecting the measurement circuit of the recorded sector $ab$. This reduction applies unchanged to the depolarized family of Eq. (S54), since

$$(1-\epsilon)\rho_{\boldsymbol{\theta}} + \epsilon\,\frac{\mathbb{I}}{4} = \frac{1}{2}\sum_{a=\pm 1}\left[(1-\epsilon)\rho_{\boldsymbol{\theta},a}^{(r)} + \epsilon\,\frac{\mathbb{I}}{2}\right] \otimes |a_x\rangle\langle a_x|.$$

The injected noise is thus an independent single-qubit depolarizing channel of strength $\epsilon$ on each $r$ qubit. We implement it by Pauli injection with derandomized weights. Each copy receives no Pauli with weight $1 - 3\epsilon/4$ and each of $\sigma_x$, $\sigma_y$, $\sigma_z$ with weight $\epsilon/4$. Shots are allocated to the Pauli combinations in exact proportion to their weights, the injections on the two copies are drawn independently, and the injection labels are withheld from the estimator. The tomography circuits of the model reconstruction omit the injection, because the ensemble-averaged channel is an exact Bloch vector contraction by $1 - \epsilon$ that is applied to the reconstructed model in calculations. Gate errors of the injection pulses therefore enter only the protocol side of the comparison. At the largest injected strength, we estimate that injection-gate errors change the normalized cost by only about $10^{-5}$. This estimate scales the single-qubit gate

error by $\epsilon$ per copy; the $\sigma_z$ injection is a virtual frame change. The estimated change is much smaller than the statistical uncertainty in the measured two-copy cost $C^{(2)}_{\rm exp}$.

The conditional unitary $U_{\boldsymbol{\theta},a}$ is a single rotation about the equatorial axis at angle $\chi_a = \mathrm{atan2}(\phi_a, \theta_1)$ from $\sigma_x$ by the angle $2\sqrt{2}\,(\theta_1^2 + \phi_a^2)^{1/2}$. When $\theta_1 = \phi_a = 0$, the rotation angle is zero and we set $\chi_a = 0$. For each sector $ab$, a basis change built from one $XX{+}YY$ interaction gate and single-qubit Cliffords maps the joint eigenbasis of the block observables of the sector to the computational basis. All settings share a single parametrized circuit structure. After transpilation to the native gate set, each protocol circuit contains two CZ gates. The data were taken on the superconducting processor `ibm_boston` with a fixed qubit pair selected from same-day calibration data (physical qubits with index 61 and 62, CZ error $9.4 \times 10^{-4}$, readout errors $1.7 \times 10^{-3}$ and $1.8 \times 10^{-3}$). We also verified that transpilation preserved the physical-qubit assignment and circuit unitary and introduced no additional two-qubit gates. We set a repetition delay of 100 $\mu$s between the final measurement of one circuit execution and the start of the next. In each run, we measure the central setting $\boldsymbol{\theta} = 0$ and six displaced settings. For each of the three parameters, we shift it by $\pm\delta$, with $\delta = 0.05$, while holding the other two at zero. We also collect readout-calibration data for the prepared computational states.

### B. Experimental measurement of two-copy cost and certified single-copy bound

The cost is estimated in two stages from independent data sets with identical settings. We use parameter shifts of $\pm\delta$, with $\delta = 0.05$, to estimate derivatives in both stages. In the design stage, we determine the empirical score estimator from the measured outcome frequencies $\hat{p}_k$ at $\boldsymbol{\theta} = 0$ and their finite-difference slopes $\partial_\nu \hat{p}_k$:

$$\widetilde{\boldsymbol{\theta}}(k) = \hat{I}^{-1}\,\frac{\partial \hat{p}_k}{\hat{p}_k}, \qquad \hat{I}_{\mu\nu} = \sum_k \frac{\partial_\mu \hat{p}_k\,\partial_\nu \hat{p}_k}{\hat{p}_k}.$$

In the evaluation stage, we estimate the response matrix $B_{\mu\nu} = \partial_\nu \langle \widetilde{\theta}_\mu \rangle$ by finite differences. Its entry $B_{\mu\nu}$ describes how the mean estimate of parameter $\mu$ changes with parameter $\nu$. We also measure the covariance matrix $V$ of $\widetilde{\boldsymbol{\theta}}$ at $\boldsymbol{\theta} = 0$ and report the calibrated cost

$$C^{(2)}_{\rm exp} = 2\,\mathrm{Tr}(B^{-1} V B^{-T}). \tag{S62}$$

With the exact local response $B$, subtracting the mean at $\boldsymbol{\theta} = 0$ and multiplying by $B^{-1}$ gives a locally unbiased estimator. In the experiment, we evaluate Eq. (S62) using the measured covariance and a response estimated from the positive and negative parameter shifts. We estimate counting uncertainties by resampling the evaluation counts while keeping the estimator determined from the design data fixed. These uncertainties do not include the error from approximating the local response with finite parameter shifts.

For the ideal model and the two-copy estimator in Sec. II D, using the response estimated at $\delta = 0.05$ gives a cost of approximately 0.6355, compared with the exact local value 5/8. This calculation describes the finite-displacement effect in the ideal model and is not applied as a correction to the hardware data.

We use state tomography to reconstruct the prepared states separately for each transmon and flag label. At each of the seven parameter settings, we measure the three Pauli components and mitigate readout errors using per-qubit calibrations from the same run. This reconstruction assumes ideal Pauli analysis rotations, ideal computational-basis states in the readout calibration, and independent readout errors on the two transmons. After readout inversion, we set negative corrected outcome probabilities to zero and renormalize them. We estimate the state derivatives from the differences between the positive and negative parameter shifts. The injected depolarization is included through the Bloch contraction described above.

The local bounds depend only on the states at $\boldsymbol{\theta} = 0$ and their first derivatives. To ensure positivity of the reconstructed states at $\boldsymbol{\theta} = 0$, we rescale any Bloch vector at this point whose length exceeds $1 - 10^{-6}$ to that length. The small positive eigenvalues introduced by this regularization are numerical parameters of the reconstruction. We then remove the derivative components parallel to the Bloch vector at $\boldsymbol{\theta} = 0$. This imposes the assumption that each conditional state's purity has no linear variation with the estimated parameters at $\boldsymbol{\theta} = 0$, as in the ideal family. The constraint prevents small radial reconstruction errors from being amplified as parameter sensitivity near a pure state. In the merged $\epsilon = 0$ data, the largest removed components are 0.070 and 0.022 for the first and second transmons, respectively. We assemble the resulting local models using Eq. (S17) and evaluate their Holevo and single-copy bounds. The reported bounds are conditional on these reconstruction and measurement assumptions.

We then derive the certified single-copy lower bound $C^{(1)}_{\rm opt}$ by minimizing the Gill–Massar relaxation for the reconstructed conditional-qubit models. For each reconstructed local model, the bound applies to any single-copy

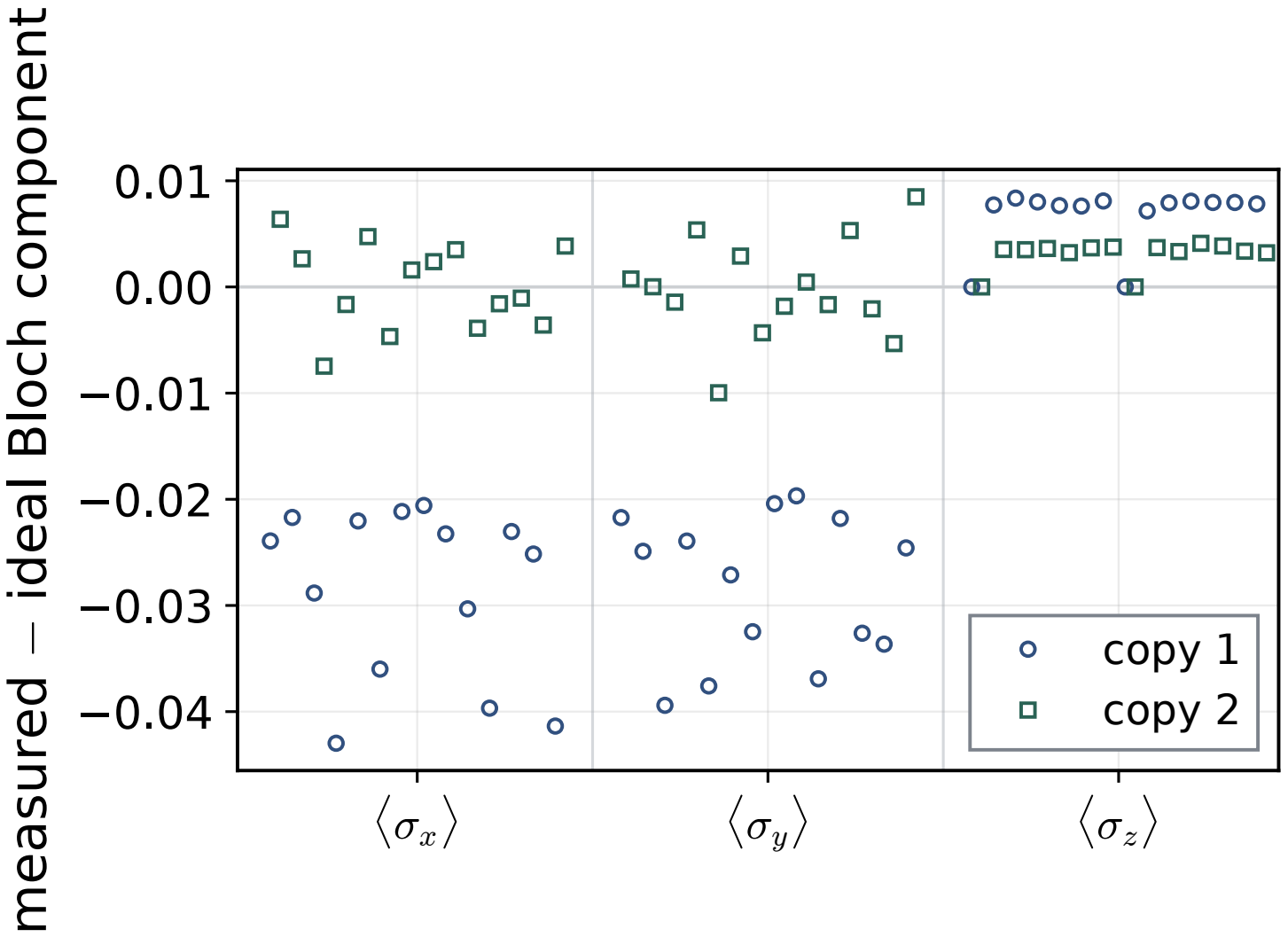


FIG. S1. Tomography reconstruction quality of the $\epsilon = 0$ data set. Residuals of the readout-mitigated conditional Bloch-vector components of both copies with respect to the ideal-model values, grouped by component, over all seven settings and both flag labels. The plotted components include readout-error correction, with negative probabilities set to zero and renormalized, before the state and derivative adjustments described in the text. The deviations are larger on the first copy.

measurement. For the ideal model, the relaxation is tight and reproduces $C^{(1)}$ exactly. We have not established tightness for the reconstructed device model, so $C^{(1)}_{\rm opt}$ is a certified lower bound rather than necessarily the exact optimum. We solve the convex minimization by Frank-Wolfe iteration. The final Frank-Wolfe gap is of order $10^{-4}$, and we subtract it from the feasible objective to obtain the certified lower bound. Shot noise in the reconstructed tangents can bias the computed limits. We estimate this bias from bootstrap resamples and use 2 times the point estimate minus the bootstrap mean, then take the smaller of the two copies' values. We estimate uncertainties as bootstrap standard deviations over resampled tomography and readout-calibration counts. The numerical certificate applies to the specified local model, and these standard deviations describe counting uncertainty within the reconstruction. They do not include systematic errors in the assumed tomography and readout model. Note that the two transmons yield slightly different reconstructed models, so we report the smaller of their per-copy bounds as $C^{(1)}_{\rm opt}$.

### C. Certified separate-measurement bound for the transmon pair

For the pair of reconstructed local models, we also derive a certified lower bound for strategies that measure the two transmons separately, allowing classical postprocessing and an adaptive choice of the second measurement. For any such strategy, the total Fisher information is $I_{r_1} + I_{r_2}$, where each term is the information contributed by each of the transmons and satisfies the Gill–Massar constraint for its reconstructed model. We minimize $2\,{\rm Tr}[(I_{r_1} + I_{r_2})^{-1}]$ over these contributions to obtain the lower bound. We use the same Frank-Wolfe structure as in the single-copy calculation, with one Gill–Massar block per copy and flag label. For identical copies, this bound reduces to the single-copy bound. From the merged $\epsilon = 0$ data, we obtain $0.778(8)$ with a Frank-Wolfe gap of $8 \times 10^{-6}$. We estimate its uncertainty from a joint bootstrap of the tomography and readout-calibration counts of both copies. The joint bootstrap mean lies below the point estimate, so we retain the unadjusted value rather than increasing it through the bias correction. We find the measured two-copy cost 5.9 standard deviations below this bound. The difference between the transmons widens this margin relative to the $C^{(1)}_{\rm opt}$ comparison because one copy is constrained by the noisier transmon.

### D. Data sets and statistical checks

We collected the $\epsilon = 0$ data in five tranches with identical circuits and layout. Across these tranches, we used 5, 10, 100, 40, and 140 thousand protocol shots per setting in each stage and 1, 2, 20, 90, and 60 thousand tomography shots

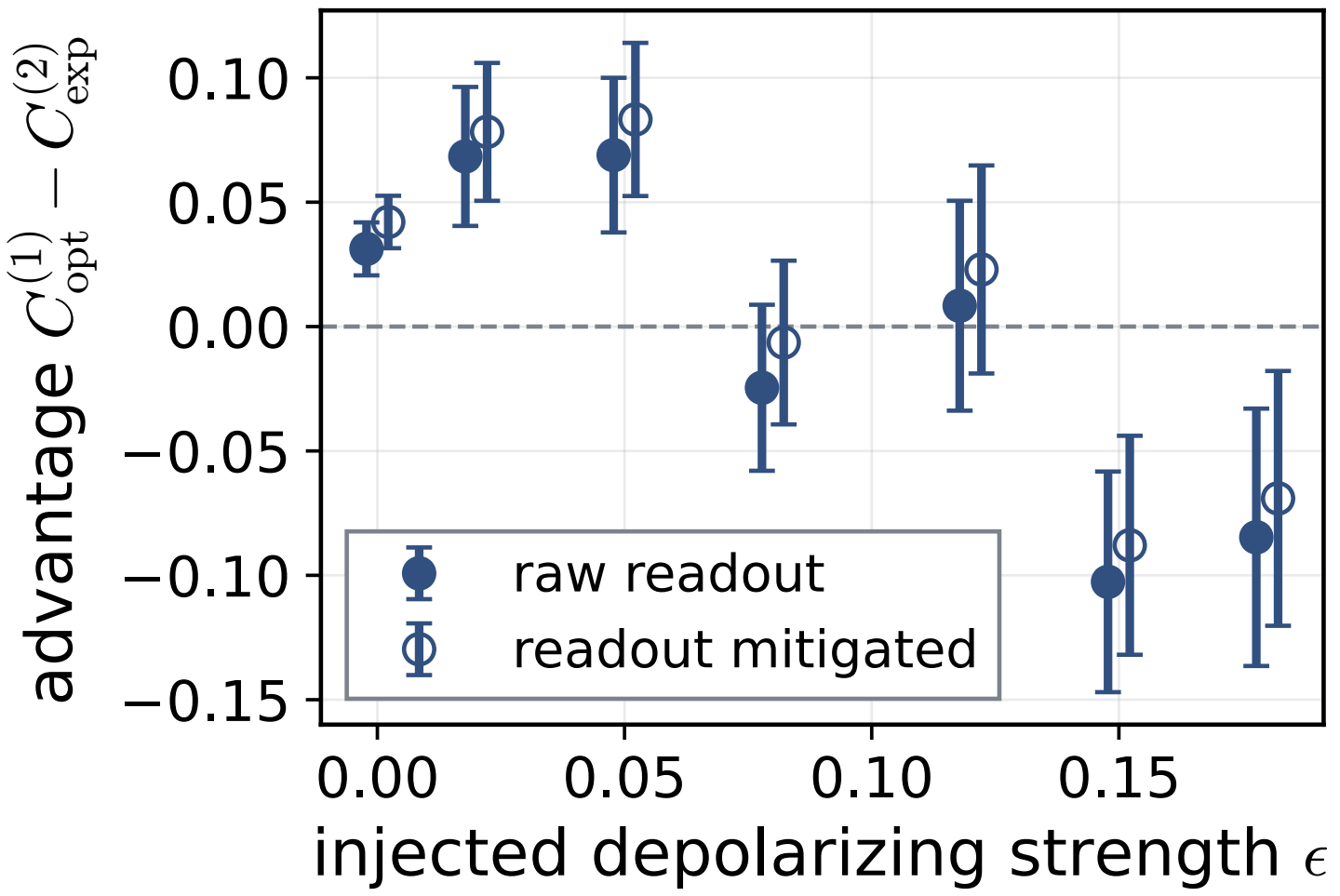


FIG. S2. Differences between the single-copy lower bound and the costs calculated from raw (filled) and readout-corrected (open) outcome probabilities across the injected-noise sweep. Error bars combine the bootstrap uncertainties of the corresponding cost and of $C^{(1)}_{\rm opt}$. The open symbols describe the inferred distribution without readout errors under the calibrated readout model.

per circuit. We varied their ratio according to the measured error scalings of the cost and of the certified single-copy lower bound $C^{(1)}_{\rm opt}$. We increased the total shot count twice while intermediate results were visible, so here we report the per-tranche costs and test their consistency before combining the counts. The five costs, 0.755(69), 0.721(46), 0.721(15), 0.669(20), and 0.715(12), give $\chi^2 = 5.3$ for four degrees of freedom ($p = 0.25$). Their weighted mean of 0.710 agrees with the merged value. We merge the tranches by adding their counts entrywise, which is statistically equivalent to a single run because each estimator in the analysis normalizes by the counts themselves. The merged totals are $4.13 \times 10^6$ protocol shots ($2.95 \times 10^5$ per setting in each stage) and $7.27 \times 10^6$ tomography shots ($1.73 \times 10^5$ per tomography circuit).

From the merged $\epsilon = 0$ data, we obtain $C^{(2)}_{\rm exp} = 0.7091(87)$ and the certified single-copy lower bound $C^{(1)}_{\rm opt} = 0.7412(74)$. Their difference gives an advantage of 0.032(11) at 2.8 standard deviations. We also find the measured two-copy cost 2.3 standard deviations below the ideal single-copy value $C^{(1)} = (3 + 2\sqrt{2})/8$. We find the measured response matrix close to the identity and show the tomography reconstruction quality in Fig. S1. For the reconstructed single-copy model of the second transmon, which is used to obtain the quoted $C^{(1)}_{\rm opt}$, we calculate a Holevo bound of 0.638. This value is evaluated using the regularized states at $\boldsymbol{\theta} = 0$ and the projected derivatives described above.

We repeat the protocol at $\epsilon \in \{0.02, 0.05, 0.08, 0.12, 0.15, 0.18\}$ with one run per point, $5 \times 10^4$ protocol shots per setting in each stage, and $3 \times 10^4$ shots per tomography circuit. We measure two-copy costs of 0.711(21), 0.762(22), 0.884(27), 0.991(33), 1.113(38), and 1.163(45), against reconstructed certified single-copy lower bounds of 0.776(21), 0.831(22), 0.861(20), 1.000(28), 1.008(22), and 1.077(24), respectively. The measured costs lie 2.2 standard deviations below $C^{(1)}_{\rm opt}$ at $\epsilon = 0.02$ and 0.05. The differences $C^{(1)}_{\rm opt} - C^{(2)}_{\rm exp}$ are consistent with zero at $\epsilon = 0.08$ and 0.12, and their point estimates are negative at $\epsilon = 0.15$ and 0.18. Across the sweep, the reconstructed $C^{(1)}_{\rm opt}$ follows the exact scaling $(1-\epsilon)^{-2}$ of the depolarized family. These comparisons do not establish where the actual advantage disappears, because $C^{(1)}_{\rm opt}$ is a lower bound whose attainability has not been established for the reconstructed models. The excess cost at $\epsilon = 0$ is consistent with an effective injected depolarizing strength of roughly 0.05 per copy.

### E. Readout error mitigation

We compute the quoted $C^{(2)}_{\rm exp}$ from protocol counts without readout mitigation, so it includes readout noise from the full measurement chain. By contrast, we mitigate the tomography data used to reconstruct $C^{(1)}_{\rm opt}$. We also use the readout calibration to infer the outcome distribution without readout errors and evaluate the corresponding cost. The sector labels are classical bits and carry no readout error. We assume independent readout errors on the two transmons and use the per-qubit readout calibrations measured in each run to mitigate the four-outcome transmon counts for every sector in both the design and evaluation stages. We draw bootstrap resamples from the raw counts,

and then mitigate each one separately to propagate the uncertainty introduced by the mitigation. Across the sweep, the cost calculated from the corrected probabilities is 0.010 to 0.018 below the raw-count value. At $\epsilon = 0$, this inferred cost is 0.698(8). The variance of an estimator applied to the measured outcomes must also include the fluctuations introduced by the inverse readout correction. Evaluating this variance for the corresponding estimator gives a cost of 0.710(8). For this variance calculation at $\epsilon = 0$, we obtain the uncertainty by resampling the protocol and readout-calibration counts together. Figure S2 compares the costs obtained from the raw and corrected outcome probabilities.

* changhaoli96@gmail.com



\end{document}